\documentclass[12pt]{article}
\usepackage{latexsym}
\usepackage{epsfig,amssymb,euscript,slashed}
\usepackage{amsmath}
\usepackage{array,calc,epsfig}
\usepackage[linktocpage]{hyperref}
\usepackage{bbm}
\usepackage{fancybox}

\def\be{\begin{equation}}
\def\ee{\end{equation}}
\def\bseq{\begin{subequations}}
\def\eseq{\end{subequations}}

\def\bea{\begin{eqnarray}}
\def\eea{\end{eqnarray}}

\def\bseq{\begin{subequations}}
\def\eseq{\end{subequations}}
\def\nn{\nonumber}

\numberwithin{equation}{section} 
\usepackage[]{graphicx}

\def\d {{\rm d}}

\def\calo         {{\cal O}}

\def\ii           {{\rm i}}

\def\sqr#1#2{{\vcenter{\vbox{\hrule height.#2pt
 \hbox{\vrule width.#2pt height#1pt \kern#1pt \vrule width.#2pt}\hrule
 height.#2pt}}}}

\def\d{\text{d}}

\def\slashchar#1{\setbox0=\hbox{$#1$}           
\dimen0=\wd0                                 
\setbox1=\hbox{/} \dimen1=\wd1               
\ifdim\dimen0>\dimen1                        
\rlap{\hbox to \dimen0{\hfil/\hfil}}      
#1                                        
\else                                        
\rlap{\hbox to \dimen1{\hfil$#1$\hfil}}   
/                                         
\fi}

\begin{document}
\font\cmss=cmss10 \font\cmsss=cmss10 at 7pt

\hfill
\vspace{18pt}
\begin{center}
{\LARGE \textbf{Brane induced supersymmetry breaking}}
\end{center}
\begin{center}
 {\LARGE \textbf{and de Sitter  supergravity}}
\end{center}


\vspace{6pt}

\begin{center}
{\textsl{\rm Igor Bandos$\,^{a,b}$, Luca Martucci$\,^{c,d}$, Dmitri Sorokin$\,^{d,c}$ and Mario Tonin$\,^{c,d}$}}

\vspace{1cm}
\textit{\small $^a$ Department of Theoretical Physics, University of the Basque Country UPV/EHU, \\ P.O. Box 644, 48080 Bilbao, Spain}\\ \vspace{6pt}
\textit{\small $^b$  IKERBASQUE, Basque Foundation for Science, 48011, Bilbao, Spain}\\ \vspace{6pt}
\textit{\small $^c$  Dipartimento di Fisica e Astronomia ``Galileo Galilei",  Universit\`a di Padova, \\ Via Marzolo 8, 35131 Padova, Italy} \\  \vspace{6pt}
\textit{\small $^d$ I.N.F.N. Sezione di Padova,
Via Marzolo 8, 35131 Padova, Italy} \\  \vspace{20pt}
\textit{E-mails}: igor.bandos@ehu.eus, luca.martucci@pd.infn.it, dmitri.sorokin@pd.infn.it, mario.tonin@pd.infn.it
\end{center}


\vspace{12pt}

\begin{center}
\textbf{Abstract}

\end{center}

\vspace{4pt} {\small
\noindent We obtain a four-dimensional supergravity with spontaneously broken supersymmetry allowing for de Sitter vacua by coupling a superspace action of minimal \hbox{$N=1$,} $D=4$ supergravity to a locally supersymmetric generalization of the Volkov-Akulov goldstino action  describing the dynamics  of a space-filling non-BPS 3-brane in $N=1$, $D=4$ superspace. To the quadratic order in the goldstino field the obtained action coincides with earlier constructions of supergravities with nilpotent superfields, while matching the higher-order contributions will require a non-linear redefinition of fields. In the unitary gauge, in which the goldstino field is set to zero, the action coincides with that of Volkov and Soroka. We also show how a nilpotency constraint on a chiral curvature superfield emerges in this formulation.

\noindent }

\vspace{1cm}


\thispagestyle{empty}


\newpage

\setcounter{footnote}{0}

%

\section{Introduction}
Recently, there has been a significant revival of interest in the role of the Volkov-Akulov goldstino \cite{Volkov:1972jx,Volkov:1973ix} in spontaneous breaking local supersymmetry and generating a positive contribution to the cosmological constant in supergravity
\cite{Dudas:2015eha,Bergshoeff:2015tra,Hasegawa:2015bza,Ferrara:2015gta,Kuzenko:2015yxa,Antoniadis:2015ala} (see \cite{Schillo:2015} for the latest review and developments)\footnote{Let us note that a super-Higgs effect via coupling the Volkov-Akulov goldstino to a simple supergravity multiplet (and to a vector gauge field) was first considered by Volkov and Soroka as early as in 1973 \cite{Volkov:1973jd,Volkov:1974ai}. The idea was to gauge Poincar\'e supersymmetry which was non-linearly realized in the Volkov-Akulov action, thus coupling the latter to gravity and the Rarita-Schwinger field. The Volkov-Soroka construction contained all the ingredients of the $N=1$, $D=4$ supergravity action, including a cosmological term and a mass term for the gravitino, but the relative coefficients between the different terms of the action  were arbitrary. As we will show in Section \ref{VS}, in the unitary gauge in which goldstino vanishes, and upon a re-scaling of the gravitino field the Volkov-Soroka action takes the same form as the `de Sitter supergravity' action. When the cosmological constant and the gravitino mass is put to zero the Volkov-Soroka action reduces to the $N=1$, $D=4$ supergravity action of \cite{Freedman:1976xh,Deser:1976eh}.}.
In these and related papers the construction is based on the constrained (nilpotent) superfield description of the Volkov-Akulov goldstino \cite{Rocek:1978nb,Lindstrom:1979kq,Casalbuoni:1988xh,Komargodski:2009rz,Komargodski:2010rb,Kuzenko:2011ti,Farakos:2013ih} (see \cite{Ivanov:1977my,Ivanov:1978mx} for the study of the general relation between linear and non-linear realizations of supersymmetry and \cite{Lindstrom:1979kq,Samuel:1982uh,Ivanov:1984hs,Ivanov:1989bh} for the extension of these methods to describe spontaneously broken supergravities in superspace). The nilpotent nature of the goldstino supermultiplet, in which the scalar fields are not elementary but composed from goldstino bilinears, has been proved important for building cosmological and inflationary models in the framework of supergravity and string theory \cite{Antoniadis:2014oya,Buchmuller:2014pla,Ferrara:2014kva,Kallosh:2014via,Dall'Agata:2014oka}.

On the other hand, metastable de Sitter vacua may arise in string theory via the KKLT construction \cite{Kachru:2003aw} involving anti-D3-branes which also break supersymmetry \`a la Volkov-Akulov \cite{Bergshoeff:2013pia,Kallosh:2014wsa,Bergshoeff:2015jxa}.
Anti-D-branes are just D-branes but with opposite Ramond-Ramond charges. So they usually completely break the supersymmetry preserved by a string compactification. In ten-dimensional supergravity regime, the dynamics of (anti) D-branes is described  by a superspace DBI-like action \cite{Cederwall:1996pv,Cederwall:1996ri,Aganagic:1996pe,Bergshoeff:1996tu}, which is manifestly invariant under local ten-dimensional supersymmetry. Hence, provided the supersymmetry breaking scales are low enough, the contribution of anti-D-branes to the four-dimensional effective theory should be naturally given by a four-dimensional superspace Green-Schwarz DBI-like action of the original Volkov-Akulov type \cite{Hughes:1986dn,Kallosh:1997aw} coupled to a superspace action describing the bulk supergravity degrees of freedom.  {This description would provide a natural alternative to the approach using nilpotent superfields and would establish a more direct link  with string theory constructions involving anti-D-branes as studied \emph{e.g.} in \cite{Ferrara:2014kva,McGuirk:2012sb,Bergshoeff:2015jxa,Kallosh:2015nia}.}

In this paper we consider a minimal possible realization of such a four-dimensional theory, namely a space-filling 3-brane carrying the Volkov-Akulov goldstino and coupled to a minimal $N=1$, $D=4$ off-shell supergravity multiplet (including auxiliary fields) formulated in curved superspace \cite{Wess:1978bu}. The system is described by the superfield action having a suggestive geometric form of the sum of three different volumes:
 \begin{eqnarray}\label{action}
S=\frac 3{4\kappa^2} \int \d^8z\, {\rm Ber} \,E + \frac m{2\kappa^2}\Big(\int \d^6\zeta_L\, {\mathcal E}+{\rm c.c.}\Big) +{f^2} \int \d^4\xi \,\det\mathbb E(z(\xi))
\; .
\end{eqnarray}
The first term corresponds to the standard pure supergravity action: $\kappa^2$ is the gravitational constant, $z^M=(x^m,\theta^{\mu},\bar\theta^{\dot\nu})$ are  coordinates of the $N=1$ $D=4$ superspace, whose  curved geometry  is described by the supervielbein $E^{A}=\d z^ME_M^A(z)$ containing the fields of the minimal off-shell supergravity multiplet \cite{Stelle:1978ye,Ferrara:1978jt,Wess:1978bu,Wess:1978ns} and ${\rm Ber} \,E$ is the usual Berezenian superspace measure (see \cite{Wess:1992cp,Buchbinder:1995uq} for the detailed description of the superspace formulation). The second term gives rise to the anti-de-Sitter cosmological term and the corresponding mass ($m$) term for the gravitino field. Here  $\mathcal E$ is the volume measure of the chiral subspace  $\zeta^{\mathcal M}_L=(x^m_L,\Theta^\mu)$.

The third term in \eqref{action} provides the full non-linear contribution  of the space-filling 3-brane to the action. It couples to  supergravity via its embedding
\be
\xi^i\mapsto z^M(\xi)=(x^m(\xi),\theta^\mu(\xi),\bar\theta^{\dot\mu}(\xi))
\ee  in the bulk superspace, where $\xi^i$  ($i=0,1,2,3$) are the brane worldvolume coordinates. $\det\mathbb E(z(\xi))$ denotes the determinant of $\mathbb E^a_i(z(\xi))\equiv \partial_i z^ME_M^a(z(\xi))$, which is the pullback of the vector supervielbein $E^a(z)=\d z^ME_M^a(z)$ on the 3-brane worldvolume. $f^2$ is the 3-brane tension which gives a positive contribution to the cosmological constant and determines the supersymmetry breaking scale\,\footnote{The coefficient in front of the 3-brane action should be positive for the $\theta(\xi)$-field  kinetic term to have a correct sign.}. The two scales, $m$ and $f$, determine the value of the cosmological constant $\Lambda=f^2-\frac{3m^2}{\kappa^2}$, which can be positive, hence allowing for the existence of de Sitter vacua in this theory. For this reason the theory under consideration was dubbed de Sitter supergravity \cite{Bergshoeff:2015tra}.
The action  (\ref{action}) thus provides its geometric formulation which is directly  related to the 3-brane realization of the Volkov-Akulov theory.

 In order to  highlight  the conceptual distinction between this approach and the ones using constrained superfields, let us note that
 the 3-brane action is invariant under worldvolume diffeomorphisms $\xi^i \rightarrow \xi'^i(\xi)$. This has the important consequence that the embedding field $x^m(\xi)$, which could be regarded as a bosonic `superpartner' of the fermionic embedding field $\theta^\mu(\xi)$, carries only pure gauge degrees of freedom which can be eliminated without any need of imposing a nilpotency constraint.
 The connection with the Volkov-Akulov theory then becomes manifest if we gauge-fix the worldvolume diffeomorphisms by imposing the static gauge $x^m(\xi)=\delta^m_i\xi^i$, i.e.\ by identifying the worldvolume coordinates with those of the  four-dimensional bulk space-time, leaving only  the Volkov-Akulov goldstino $\theta^\mu(x)$ as the physical worldvolume field. Indeed, in the flat space limit this term of the action \eqref{action} reduces to the original Volkov-Akulov action \cite{Volkov:1972jx,Volkov:1973ix}.

Note that the 3-brane action does not have a Wess-Zumino term (simply because it does not exist in the minimal model under consideration) and does not possess kappa-symmetry, which are intrinsic ingredients of the 1/2 BPS superbranes preserving, at least locally, half of bulk supersymmetry\footnote{The interaction of minimal Einstein supergravity with ${N}=1$, $D=4$  BPS branes (massless superparticle, superstring and supermembrane) was studied in \cite{Bandos:2002bx,Bandos:2003zk,Bandos:2012gz}. }. Hence, in our case the whole $N=1$ bulk supersymmetry is spontaneously broken. In other words, the action \eqref{action} is manifestly invariant under the superdiffeomorphisms $\delta z^M(z)$ which incorporate the local supersymmetry transformations. Under such superdiffeomorphisms the embedding coordinates $z^M(\xi)$ must transform accordingly, $z^M(\xi)\rightarrow z^M(\xi)+ \delta z^M(z(\xi))$, hence leading to the presence of the goldstino  which
(upon imposing the static gauge)
 transforms under supersymmetry in a non-linear way. This provides a simple low-energy realization of the same supersymmetry breaking mechanism generated by introducing anti-D-branes in string compactifications.

In the rest of the paper we will study the component field structure of the action \eqref{action}. As we will see, the interaction of the goldstino fields with the supergravity  multiplet is encoded in the 3-brane  action via the dependence of the (pull-back of) the vector supervielbein $E^a(x,\theta,\bar\theta)$ on the supergravity fields whose explicit form to all orders in $\vartheta$ we will derive in Section \ref{expansion}. In particular, the coupling of the 3-brane to a complex scalar auxiliary field of the ``old minimal" supergravity produces a solution (in terms of the goldstino) of the nilpotency constraint on a chiral scalar curvature superfield $R(z)$ similar to that used in \cite{Antoniadis:2014oya,Dudas:2015eha} to construct nilpotent supergravity models.   We will then compare our action with the form of the dS supergravity action constructed with the use of the nilpotent goldstino superfield \cite{Bergshoeff:2015tra,Hasegawa:2015bza}, and with the Volkov-Soroka model \cite{Volkov:1973jd,Volkov:1974ai}.

\setcounter{equation}0
\section{Component form of the $N=1$ supergravity action coupled to the 3-brane}

The main result of this paper is the derivation of the component form of the action (\ref{action}). It is obtained by integrating the first two terms of (\ref{action}) over the Grassmann-odd coordinates and fixing the static-gauge $x^m(\xi)=\delta^m_i\xi^i$ on the 3-brane, as described in the Introduction. We thus get the following action\footnote{\label{resc}To bring the normalization of the fermionic kinetic and mass terms in this action to a canonical form one should re--scale the gravitino and the goldstino fields as follows
$\psi \rightarrow \frac 12 \psi$, $\theta\rightarrow \frac 1f\theta$ and $\bar\theta\rightarrow \frac 1f\bar\theta$.} (see Appendices \ref{notation} and \ref{constraints} for our notation and conventions):
\be\label{adssugra+3b}
\begin{aligned}
S=&S_{\rm SG}+S_{\text{VA}}\quad~\text{with}\\
S_{\rm SG}=&\frac 1{2\kappa^2}\int \d^4x\, e\Big[\mathcal R(\hat\omega) - 4 e^{-1}\varepsilon^{mnkl} (\hat\nabla_n\psi_{k}\sigma_{l} \bar{\psi}_m +{\psi}_m \sigma_n\hat\nabla_k\bar\psi_l)\\ &- 4m(\bar{\psi}{}^a\sigma_{ab}\bar{\psi}{}^b+{\psi}{}^a\sigma_{ab}{\psi}{}^b) + {3\over 32}  G_aG^a+\frac 38(4m +R)(4m +{\bar R})-{6m^2} \Big]\\
S_{\text{VA}}=&{f^2} \int \d^4x\, \det\mathbb E(x,\theta(x),\bar\theta(x))\,.
\end{aligned}
\ee
$S_{\rm SG}$ gives the standard pure $N=1$ AdS supergravity: $e=\det e^a_m(x)$, where $e^a_m(x)$ is the space-time vielbein, $\psi_m^\alpha(x)$ and $\bar\psi_m^{\dot\alpha}(x)$ are the Weyl-spinor gravitino field and its complex conjugate,
the covariant derivative $\hat\nabla=\d-\hat\omega$ is defined in Appendix \ref{constraints} -- see eqs. \eqref{conomega}--\eqref{nabla} -- and $\mathcal R(\hat\omega)$ is the curvature scalar associated with $\hat\omega$. If in \eqref{adssugra+3b} the connection $\hat\omega^{ab}$ gets substituted with its expression in terms of the ordinary spin connection $\omega^{ab}$ and gravitino bilinears, the action exhibits quartic gravitino terms. Finally, $G^a$,  $R$ and $\bar R=(R)^*$ are the old minimal supergravity auxiliary fields. When $m=0$, $S_{\rm SG}$ reduces to the old minimal off-shell supergravity action derived in \cite{Stelle:1978ye,Ferrara:1978jt}.

On the other hand, the coupling of the Volkov-Akulov goldstino to the supergravity fields is encoded in $S_{\text{VA}}$, in which
 $\det\mathbb E$ denotes the determinant of the worldvolume pullback of the bulk vector supervielbein
\bea
\mathbb E_m^a(x,\theta(x),\bar\theta(x))&=&E^a_m(x,\theta(x),\bar\theta(x))+\partial_m\theta^\alpha(x) E^a_\alpha(x,\theta(x),\bar\theta(x))\\
&+&\partial_m\bar\theta^{\dot\alpha}(x)E^a_{\dot\alpha}(x,\theta(x),\bar\theta(x))\,.\nonumber
\eea
Here $\theta^\alpha(x)$ and $\bar\theta^{\dot\alpha}(x)$ are the components of the Volkov-Akulov goldstino and
the complicated non-linear structure of $S_{\text{VA}}$ is encoded in the $\theta$-expansion of $\mathbb E_m^a(x,\theta(x),\bar\theta(x))$.  Its explicit form will be discussed in detail in Section \ref{expansion}, but we anticipate some
of its implications.

In the action \eqref{adssugra+3b} the auxiliary fields $R$ and $G^a$ can be expressed in terms of the physical fields by solving their equations of motion
\begin{subequations}\label{eomaux}
\begin{align}
\label{eomaux1}
R&=-4m-\frac {16\kappa^2f^2}{3} \,  \left(\frac {\delta \mathbb E_m^a}{\delta{\bar
R}}\,\mathbb E^m_a\right)\,\frac{{\det }\,\mathbb E}e+\frac {16\kappa^2 f^2}{3e} \,
\partial_n\left(\frac {\delta \,{\det}\, \mathbb E}{\delta({\partial_n\bar
R})}\right)\,,\\
\label{eomauxG1}
G_b&=- \frac {32\kappa^2f^2}{3}  \, \left(\frac {\delta
\mathbb E_m^a}{\delta{G^b}}\,\mathbb E^m_a\right)\,\frac{{\det }\,\mathbb E}e+\frac {32\kappa^2f^2}{3e}
\,\partial_n\left(\frac {\delta \,{\det}\, \mathbb E}{\delta({\partial_n
G^b})}\right),
\end{align}
\end{subequations}
where $\mathbb E^m_a$ is inverse of $\mathbb E^a_m$. As we will see, the auxiliary fields appear in $E^a(x,\theta,\bar\theta)$ starting from the second order at most in quadratic combinations and may only appear linearly under the space-time derivative at the fourth order in $\theta,\bar\theta$. So \eqref{eomaux} are algebraic linear equations for $R$ and $G^a$,  and are thus exactly solvable. In other words, the auxiliary fields can be integrated out in \eqref{adssugra+3b} by the standard Gaussian integration.

The bulk supergravity action $S_{\rm SG}$ is invariant under the following local supersymmetry transformations of the supergravity fields with parameter $\epsilon^\alpha(x)$, see e.g. \cite{Wess:1992cp,Buchbinder:1995uq}:
\be\label{susy}
\begin{aligned}
\delta e^a_m=&2\ii(\epsilon\sigma^a\bar\psi_m-\psi_m\sigma^a\bar\epsilon)\;,\\
\delta\psi_m=&\hat \nabla \epsilon+\frac{\ii}{8}\,\bar R(\bar\epsilon\tilde\sigma_m)^\alpha+
\frac{\ii}{16}\left(3G_m \epsilon^\alpha -(\epsilon\sigma_a\tilde{\sigma}_m)^\alpha G^a  \right)\;,\\
 \delta\bar\psi_m=& \hat \nabla \bar\epsilon+\frac{\ii}{8}\,\bar R(\tilde\sigma_m\epsilon)^{\dot\alpha}-\frac{\ii}{16}\left(3G_m \bar{\epsilon}^{\dot\alpha} -(\tilde{\sigma}_m\sigma_a \bar{\epsilon})^{\dot\alpha} G^a  \right)\;, \\
\delta R  =&- \frac{16}{3}\hat{\nabla}_m\psi_n{\sigma}^{mn}\epsilon
 -2\ii \epsilon\sigma_a  \bar{\psi}^a\, R - \ii \epsilon\psi_a\, G^a\;,\\
 \delta \bar R =& - \frac{16}{3} \bar\epsilon\tilde\sigma^{mn}\hat{\nabla}_m\bar{\psi}_n
 +2\ii\psi^a\sigma_a \bar{\epsilon}\, \bar R +\ii \bar\psi_a\bar\epsilon\, G^a\;,\\
  \delta G_a =& - \frac{40\ii}{3e} e_{m a}\varepsilon^{mnkl} \big(\hat{\nabla}_n\psi_k \sigma_l \bar{\epsilon} -{\epsilon}\sigma_l   \hat{\nabla}_n\bar{\psi}_k\big)- \frac{32}{3}  e^{[m}_{ a}\big(\hat{\nabla}_{m}\psi_{n }{\sigma}^{n]}\bar\epsilon +\epsilon {\sigma}^{n]}\hat{\nabla}_{m}\bar\psi_{n}\big)   \\ & +  \ii G^a\big(\psi_b\sigma^b\bar{\epsilon}- {\epsilon}\sigma^b\bar\psi_b\big) + {\ii\over 2}\epsilon_{abcd} \big(\psi^b\sigma^c\bar{\epsilon}+\varepsilon\sigma^b\bar\psi^c\big)G^d - 2\ii R \bar{\psi}_a\bar{\epsilon}+2\ii\bar{R}
\epsilon{\psi}_a\;.
\end{aligned}
\ee
On the other hand, $S_{\rm VA}$  is invariant under the combined action of (\ref{susy}) and the following supersymmetry variations of the goldstino
\be\label{susyth}
\begin{aligned}
\delta\theta^{\alpha}(x)&=-\epsilon^\alpha(x,\theta(x),\bar\theta(x))+\epsilon^m(x,\theta(x),\bar\theta(x))\partial_m\theta^\alpha(x) \,,\\
\end{aligned}
\ee
where the term with the derivatives of $\theta(x)$ appears in \eqref{susyth} because of a contribution of a worldvolume diffeomorphism required to preserve the static gauge  $x^m(\xi)=\delta^m_i\xi^i$. The form of $\epsilon^\alpha(x,\theta,\bar\theta)$ and $\epsilon^m(x,\theta,\bar\theta)$
is determined from the requirement of the preservation of the Wess-Zumino gauge with the use of the procedure explained in Appendix \ref{WZg}. To the second order in $\theta,\bar\theta$ we thus get
\be\label{varTh}
 \begin{aligned}
 \delta \theta^{\alpha}=&- \epsilon^\alpha    -\ii\left(\theta \sigma^m\bar\epsilon-  \epsilon\sigma^m\bar\theta \right)\; \left[\psi_m^\alpha +\nabla_m \theta{}^\alpha
 -\ii \left( \theta\sigma^n\bar\psi_m- \psi_m \sigma^n\bar\theta \right)(\psi_n^\alpha+\nabla_m \theta{}^\alpha)
 \right]\\ & +
 \frac 1{16} \left(\theta \sigma^a\bar\epsilon-  \epsilon\sigma^a\bar\theta \right)\; \left[2
 \theta^\alpha G_a + (\theta \sigma_{ab})^{\alpha} G^b+2 (\bar\theta\tilde\sigma_a)^{\alpha} R\right] +  \ldots\, , \qquad
\end{aligned}
\ee
which explicitly shows that $\theta^\alpha(x)$ is a goldstino field which gets shifted and non-linearly transformed under supersymmetry, implying the spontaneous breaking of the latter.

In the rest of this section we discuss in more  detail the derivation and the structure of this action.

\subsection{$N=1$, $D=4$ AdS supergravity action}
Let us start with a brief review of a derivation of the action $S_{\rm SG}$ in (\ref{adssugra+3b}). This corresponds to the bulk superspace contribution to (\ref{action}):
\be\label{action1}
S_{\rm SG}=\frac 3{4\kappa^2} \int \d^8 z\, {\rm Ber}\,E + \frac m{2\kappa^2} \Big(\int \d^6\zeta_L\, {\mathcal E}+\text{c.c.}\Big)
\; .
\ee
It is well known (see e.g. \cite{Wess:1992cp}) that in order to express (\ref{action1}) in terms of the component fields of the minimal off-shell supergravity
one must use the supergravity constraints (see Appendix \ref{constraints}) and impose the Wess-Zumino gauge, which fixes part of the superdiffeomorphisms so that what remains is local supersymmetry and four-dimensional diffeomorphisms. The Wess-Zumino gauge can be written in the form
\be\label{WZgauge}
\iota_\vartheta E^A (z):= \vartheta^{\hat{\mu}}  E_{\hat\mu}{}^A(z)= \
\vartheta^{\hat{\mu}} \delta_{\hat{\mu}} {}^A\;      , \qquad \iota_\vartheta \Omega^{ab}(z):= \vartheta^{\hat\mu}  \Omega_{\hat\mu}{}^{ab}(z)= 0
 \;, \qquad
\ee
where we have introduced the 4-component fermionic coordinates $\vartheta^{\hat\mu}=(\theta^\mu,\bar\theta^{\dot\mu})$  and $\Omega^{ab}(z)=\d z^M\Omega_M{}^{ab}(z)$ is a Lorentz-algebra valued spin connection superform\footnote{{The complete set of gauge fixing conditions of the Wess--Zumino gauge  in the form of equations (\ref{WZgauge}) can be found in \cite{Bandos:1985un,Bandos:1985phd} as well as in easier accessible \cite{Bandos:2002bx}.}}. Note that in the Wess-Zumino gauge
 the indices of curved superspace fermionic coordinates
 $\hat{\mu}= (\mu,\dot\mu)$ get converted into the $SL(2,C)$ spinor indices $(\alpha,\dot\alpha)$ which we sometimes collect in 4-valued $\hat{\alpha}$.

The fields of  minimal $N=1$ supergravity are the lowest $(\theta=\bar\theta=0)$ components of the supervielbein and of the superfields $R(z)$, $\bar R(z)$ and $G^a(z)$ appearing in the expressions for the superspace torsion and curvature (see Appendix \ref{constraints})
\be\label{components}
\begin{aligned}
&e^a_m (x)=E^a_m(x,0), \quad \psi^{\alpha}_m(x)= E^\alpha_m(x,0),\quad \bar\psi^{\dot\alpha}_m(x)= \bar E^{\dot\alpha}_m(x,0), &
\\
&\quad R(x)=R(x,0), \quad \bar R(x)=\bar R(x,0), \quad G^a(x)=G^a(x,0)\,.&
\end{aligned}
\ee
Notice that we  denote the auxiliary fields $R(x)$, $\bar R(x)$ and $G^a(x)$ with the same letters as their superfield counterparts.

Due to our choice of the torsion constraint, see Eq.~\eqref{4WTa=}, the conventional supergravity connection $\hat \omega^{ab}(x)$ is related to the lowest component of ${\Omega}_{m}{}^{ab}$ as follows
\be\label{omega}
\hat\omega_{m }^{ab}:= \omega_{m }^{ab}+2\ii(\psi^{[a}\sigma^{b]}\bar{\psi}_m+\psi_{m}\sigma^{[a}\bar{\psi}^{b]}+\psi^{[a}\sigma_{m}
   \bar{\psi}^{b]})={\Omega}_{m}{}^{ab}\vert_0+{1\over 8} e_{m\, c}\varepsilon^{abcd} G_d\vert_0,
\ee
where $\omega_{m }^{ab}(x)$ is the standard (torsion-less) spin connection expressed in terms of the vielbein $e^a_m(x)$ and $\vert_0:=\vert_{\theta=0}$ stands for the lowest component of the superfields such as $\Omega^{ab}$ and $G_d$. In what follows, to make expressions shorter, we will use the connection $\hat\omega^{ab}$.

In the Wess-Zumino gauge the chiral measure $\mathcal E$ has the following form
\begin{eqnarray} \label{cE=e+}
 {\cal E} = e\Big[ 1 + 2\ii\Theta^\alpha (\sigma_a\bar{\psi}^a)_\alpha +\Theta\Theta \Big(  {3\over 4} \bar{R} - 2\bar{\psi}^a\sigma_{ab}\bar{\psi}^b \Big)\Big] \, ,
\end{eqnarray}
where $\Theta$ is a `new Grassmann coordinate' defined in \cite{Wess:1992cp} and references therein.\footnote{ $\Theta$ variable is defined \cite{Wess:1992cp} as a function of $\theta$ and other superspace coordinates ($x$ and $\bar{\theta}$) by the requirement that the coefficients in the decomposition of the covariantly chiral superfield ($\bar{{\cal D}}_{\dot{\alpha}} \Phi=0$) are given by the leading components of the covariant Grassmann derivatives ${\cal D}_\alpha\Phi\vert_0$ and $-\frac 12  {\cal D}^\alpha{\cal D}_\alpha\Phi\vert_0$, namely, $\Phi =\Phi\vert_0 + \Theta^\alpha ({\cal D}\Phi\vert_0) + {1\over 2} \Theta\Theta (-\frac 12 {\cal D}^\alpha{\cal D}_\alpha\Phi\vert_0)$. Then the chiral measure contains the standard Berezin integration with respect to  $\Theta$, $\int \d^6\zeta_L ... = \int \d^4x {\partial\over \partial\Theta^1 }{\partial\over \partial\Theta^2} =\frac 12  \int \d^4x \epsilon^{\alpha\beta}{\partial\over \partial\Theta^\alpha }{\partial\over \partial\Theta^\beta}
$. Notice that the volume of chiral superspace can be also written with the use of the complete superspace measure $\int \d^6\zeta_L {\cal E}= 2 \int \d^8 z \, \frac {{\rm Ber}\,  E}{R}$, but with the $R$ superfield in the denominator.}

So, integrating \eqref{action1} over the fermionic coordinates we get the component field action $S_{\rm SG}$ appearing in (\ref{adssugra+3b}). If the complete action were just given by  $S_{\rm SG}$,  one could integrate-out the auxiliary fields by substituting into $S_{\rm SG}$ the solution of the associated equations of motion
\be
\label{eoma}
G^a=0,\qquad R=\bar R=-4m\,.
\ee
This produces the $N=1$ supergravity action with negative cosmological constant  $\Lambda=- \frac{3m^2}{\kappa^2}$  and  gravitino mass $m$ constructed in  \cite{Townsend:1977qa}
\be\label{adssugra1}
\begin{aligned}
S_{\rm AdS}&= \frac 1{2\kappa^2}\int \d^4x\, e\Big[\mathcal R(\hat\omega) - 4 e^{-1}
\varepsilon^{mnkl} (\hat\nabla_n\psi_{k}\sigma_{l} \bar{\psi}_m +{\psi}_m \sigma_n\hat\nabla_k\bar\psi_l)\\
&\quad~~~~~~~~~~~~~~~~~  - 4m(\bar{\psi}{}^a\sigma_{ab}\bar{\psi}{}^b+{\psi}{}^a\sigma_{ab}{\psi}{}^b) -6m^2\Big]\,.
\end{aligned}
\ee
In the next subsection we will discuss how this conclusion is modified by the presence of the 3-brane.

\subsection{Coupling the 3-brane to supergravity}\label{expansion}
When we consider the coupling of supergravity to the 3-brane as in \eqref{action}, the 3-brane back-reacts, i.e. contributes to the supergravity equations of motion and, in particular, to those of the auxiliary fields modifying them. Therefore,  we cannot use the solutions \eqref{eoma} anymore and we should find their modification due to the presence of the 3-brane.

As already mentioned, after fixing the worldvolume diffeomorphism invariance by imposing the static gauge $x^m(\xi)=\delta^m_i\xi^i$, the interaction of the Volkov-Akulov goldstino with the supergravity multiplet is encoded in the 3-brane action
\be\label{3baction}
S_{VA}=f^2\int \d^4x\, {\rm det}\, \mathbb E_m^a(x,\theta(x),\bar\theta(x))
\ee
 via a complicated dependence of the supervielbein $E^a(x,\theta,\bar\theta)$ on the supergravity component fields. To get the explicit form of this dependence we should expand $E^a(x,\theta,\bar\theta)$ in powers of $\theta$ and $\bar\theta$. The series stops at the fifth order in the fermions (including $\d\theta^\alpha$ and $\d\bar\theta^{\dot\alpha}$).

To compute the $\theta$-expansion of $E^A(x,\theta,\bar\theta)$ in the Wess-Zumino gauge \eqref{WZgauge} we use the following well known procedure\footnote{Alternatively, one might adapt to our case the general expressions given in \cite{Grisaru:2000ij} (see Appendix \ref{Majorana}). { Another alternative procedure to arrive at recurrent relations, described in \cite{Bandos:2002bx} (going back to \cite{Bandos:1985un} and \cite{Bandos:1985phd}), uses the operator $\vartheta^{\hat\alpha} \partial_{\hat\alpha}=\vartheta^{\hat\alpha} {\mathcal D}_{\hat\alpha}$ instead of $\frac {\d}{\d t}$ applied to superfields with  $\vartheta\mapsto t\vartheta$.}}. Let us again use the four-component fermionic coordinates  $\vartheta^{\hat\mu}=(\theta^\mu,\bar\theta^{\dot\mu})$  introduced in the previous section.  We can then rescale them, $\vartheta \rightarrow t\vartheta$, and define the $t$-rescaled supervielbein
\be\label{t-exp}
\begin{aligned}
E^A(t):=&E^{A}(x,t{\vartheta})= \d x^m E_m^A(x, t\vartheta) + t\d\vartheta^{{\hat{\alpha}}} E_{{\hat{\alpha}}} {}^A (x, t\vartheta)\\ =&
E^A_{(0)}+tE^A_{(1)}+t^2 E^A_{(2)}+t^3 E^A_{(3)}+t^4 E^A_{(4)} +t^5 E^A_{(5)} \; ,
\end{aligned}
\ee
where  $E^A_{(n)}$ (with $n=0,\ldots, 5$) stands for the term of order $\vartheta^n$ in $E^A(x,\vartheta)$ when $t=1$.
This trick allows us to identify the different components of the $\vartheta$-expansion of $E^A$ by finding \eqref{t-exp} order by order in $t$.

To this end, first of all we observe that, taking into account the Wess-Zumino gauge \eqref{WZgauge}, we have
(see Appendix \ref{constraints} for a summary of the superspace geometry needed in the following)\begin{subequations}
\begin{align}
\frac \d{\d t}E^A(t)&=\mathcal D\iota_\vartheta E^A(t)+\iota_\vartheta T^A(t), \label{dt}\\
\label{dw}
\frac \d{\d t}{\Omega}^{ab}(t)&=\iota_\vartheta \mathcal R^{ab}(t)\,.
\end{align}
\end{subequations}
Hence taking into account the supergravity constraints we get a system of equations which can be solved order by order in $t$
\begin{subequations}\label{dsE}
\begin{align}
\label{dea}
\frac \d{\d t}E^a&=\iota_\theta T^A=2\ii\theta\sigma^a\bar E-2\ii E\sigma^a\bar\theta\, ,\\
\label{dealpha}
\frac \d{\d t}E^\alpha&=\mathcal D\theta^\alpha+\iota_\theta T^\alpha=\mathcal D\theta^{\alpha}+\frac \ii 8 E^c[(\theta\sigma_c\tilde \sigma_d)^\alpha G^d+\bar \theta_{\dot\beta}\tilde\sigma_c^{\dot\beta\alpha}  R]\, ,\\
\label{debalpha}
\frac \d{\d t}E^{\dot\alpha}&=\mathcal D\bar\theta^{\dot\alpha}+\iota_\theta T^{\dot\alpha}=\mathcal D\bar\theta^{\dot\alpha}-\frac \ii8 E^c[(\tilde \sigma_d\sigma_c\bar\theta)^{\dot\alpha} G^d+ \tilde\sigma_c^{\dot\alpha\beta}\theta_{\beta}\bar R]\,,\\
\label{dw=4D}
\frac \d{\d t} {\Omega}^{\alpha\beta}&= -  E^{(\alpha } \theta^{\beta
)}\bar{R} - \frac \ii8 E^c \left(
\tilde{\sigma}_c^{\dot{\gamma}(\alpha }\theta^{\beta )}\bar{{\cal
D}}_{\dot{\gamma}}\bar{R} - (\theta\sigma_c \tilde{\sigma}_d)^{(\alpha
} {\cal D}^{\beta )}G^d+ (\sigma_c\bar{\theta})_\gamma
W^{\alpha\beta\gamma} \right)\, ,
\end{align}
\end{subequations}
where we are implicitly using $t$-rescaled superfields.

For our purposes, we are interested in computing the explicit form of the vector supervielbein $E^a(x,\theta,\bar\theta)$.  Eq. \eqref{dea} allows us to  express $E^a(x,\theta,\bar\theta)$ in terms of the $\vartheta$-expansion of the fermionic vielbeins $E^\alpha$ and $\bar E^{\dot\alpha}$, namely,
\be\label{Ea}
\begin{aligned}
E^a=&e^a+E^a_{(1)}+E^a_{(2)}+E^a_{(3)}+E^a_{(4)}+E^a_{(5)}\\
=&\d x^me_m^a
-2\Big[\ii\Big(\psi+\frac 12  E_{(1)}+\frac 13 E_{(2)}+\frac 14  E_{(3)}+\frac 15  E_{(4)}\Big)\sigma^a\bar\theta\,+ \,{\rm c.c.}\Big]\,.
\end{aligned}
\ee
On the other hand, one can use (\ref{dealpha}) and (\ref{debalpha}) to identify the first term in the $\vartheta$-expansion of the spinorial supervielbein:
\be\label{se2}
E^\alpha= \psi^\alpha +E^\alpha_{(1)} = \psi^\alpha  +  \hat\nabla\theta^\alpha + {\ii\over 16} e^b \left[2\theta^\alpha G_b +
(\theta\sigma_{[b}\tilde{\sigma}_{c]})^\alpha G^c\right] + {\ii\over 8} e^b
(\bar{\theta}\tilde{\sigma}_b)^\alpha {R} + {\cal O}( {\vartheta}^2)
\; .
\ee
By plugging \eqref{se2} back into \eqref{Ea} we can  immediately read off the explicit expansion of $E^a$ up to the second order in $\vartheta$:
\be\label{Ea=WZ2}
\begin{aligned}
E^a =& e^a + 2\ii\theta\sigma^a\bar{\psi} - 2\ii\psi\sigma^a\bar{\theta}  +  \ii\theta\sigma^a\hat\nabla\bar{\theta} -  \ii\hat\nabla\theta\sigma^a\bar{\theta}   \\ &  + {1\over 4}e^bG_b\theta\sigma^a\bar{\theta}  - {1\over 4}e^{[a}G^{b]}\theta\sigma_b\bar{\theta}  + {1\over 8}e^a \, (\theta\theta \bar{R} + \bar{\theta}\bar{\theta} R)+
{\cal O}( \vartheta^3)\;.
\end{aligned}
\ee
Note that when the auxiliary fields are put to zero, the terms entering the second order expansions \eqref{se2} and \eqref{Ea=WZ2} coincide with the supervielbeins first constructed in \cite{Volkov:1973jd} (see Section \ref{VS}).

One can then iterate the above procedure to identify the higher order terms. At the third order we obtain
\be\label{3rdEb=Ig3}
\begin{aligned}
E^{a}_{(3)}
=&
e^b  \bar{\theta} \bar{\theta}\, \Big[ - \frac \ii6  \theta\sigma^a\bar\psi_b\, R  - \frac \ii6 \theta\psi_{b} \, G^a-\frac \ii6 \theta\sigma^a\tilde\sigma_c\psi_{b}\,  G^c \\
&\quad~~~~  + \frac \ii{3}\theta\sigma^a {\bar{\mathcal T}}_b+ \frac \ii{6}\theta\sigma_b{\bar{\mathcal T}}^a- \frac \ii{12}\theta\sigma^a{}_b\sigma_c{\bar{\mathcal T}}^c    - \frac {2}{9} \eta^{ad}(\theta \sigma_{[d}{}^c{}\mathcal T_{b]c}) \\ &  \quad~~~~ - \frac {2}{9} \Big( \eta^{ac}\theta \mathcal T_{cb}  +\frac\ii2 \eta_{be}\varepsilon^{aecd} \theta \mathcal T_{cd} \Big)
\Big]\,\,+\,\,\text{c.c.}\,,
\end{aligned}
\ee
where we have introduced\footnote{Note that the $R$-term and a part of contributions proportional to $G^a$ of \eqref{Tabal:=} do not enter the expression in the last line of  (\ref{3rdEb=Ig3}) so that instead of (\ref{Tabal:=}) one can substitute there a shorter expression
$\mathcal T_{ab}{}^\alpha\mapsto \psi_{ab}^\alpha  - {3\ii\over 8} \psi_{[a}{}^{{\alpha}}G_{b]}$. Note also that when for the pure supergravity the equations of motion for the auxiliary fields are solved as in \eqref{eoma}, $\bar{\mathcal T}^a_{{\dot\alpha}}(x)$ is proportional to the gravitino field equations.}
\begin{subequations}
\begin{align}
\label{psiab}
\psi_{ab}^\alpha:=&2e_a^me_b^n\hat\nabla_{[m}\psi^\alpha_{n]}\,\\
\label{Tabal:=}
\mathcal T_{ab}{}^\alpha:=&T_{ab}{}^\alpha\vert_0= \psi_{ab}^\alpha -{i\over 8} (\psi_{[a}\sigma_{b]}\tilde\sigma^{c})^{\alpha}G_c - {\ii\over 8} \psi_{[a}{}^{{\alpha}}G_{b]} - {\ii\over 4}
(\bar\psi_{[a}\tilde\sigma_{b]})^{\alpha}R\\
\label{RSEq:=}
\bar{\mathcal T}^a_{\dot{\alpha}} :=&\varepsilon^{abcd} \mathcal T_{bc}{}^\alpha\sigma_{d\alpha\dot{\alpha}} = \varepsilon^{abcd} \psi_{bc}{}^\alpha\sigma_{d\alpha\dot{\alpha}}+ {1\over2} (\bar\psi_b\tilde\sigma^{ab})_{\dot{\alpha}}R +{\ii\over 8} \varepsilon^{abcd}  (\psi_b\sigma_c)_{\dot{\alpha}}G_d +{1\over 2} (\psi_c\sigma^{[c})_{\dot{\alpha}}G^{a]}\,.
\end{align}
\end{subequations}
The forth and fifth order contributions to the $\vartheta$-expansion of $E^a$ are
\begin{subequations}
\begin{align}
\label{4thEb=Ig}
E^a_{(4)} =& -\frac {\ii}{2}E_{(3)}\sigma^a\bar\theta  +\frac {\ii}{2} \theta\sigma^a\bar E_{(3)} =\nn\\
=&  \;  \frac \ii{24}\, \theta\theta \, \hat\nabla\theta\sigma^a\bar\theta\, \bar R
   -\frac \ii{24}\, \bar \theta\bar \theta \,\left( \hat\nabla\theta^\gamma\,\theta_\gamma G^a   + \hat\nabla\theta\sigma^c\tilde\sigma^a\theta\,  G_c  \right)  \nn\\
 & - \frac \ii{24}\, \bar\theta\bar\theta \, \theta\sigma^a\hat\nabla\bar\theta\, R
       +\frac \ii{24}\, \theta\theta \, \left( \hat\nabla\bar\theta_{\dot{\gamma}}\,\bar\theta{}^{\dot{\gamma}}\, G^a + \hat\nabla\bar\theta\tilde\sigma^c \sigma^a\bar\theta\, G_c \right)   \nn \\
& + \, \theta\theta \, \bar \theta\bar \theta \,  e^a \Big[  \frac 1{384}  \left(  {\cal D}{\cal D} R + \bar{{\cal D}}\bar{{\cal D}}\bar R  \right)\vert_0  +
 \frac 1{192}G_cG^c-\frac 1{96}  R\bar R    \Big]  \\
&   + \, \theta\theta \, \bar \theta\bar \theta \,  e^b \Big[ \frac 1{768} \left(4\eta^{ac}\eta_{bd} +  6\delta_{[b}{}^a\delta_{d]}{}^c\right) (\tilde\sigma_c )^{\dot{\alpha}\alpha}[  \bar{\cal D}_{\dot\alpha}, {\cal D}_\alpha   ] G^d\vert_0     \nn  \\
&  + \frac 1{64} \varepsilon^{a}{}_{bcd} {\cal D}^c G^d \vert_0  - \frac {5}{192}  G_bG^a   +\frac \ii{12}\psi^\alpha  {\cal D}_\alpha G^a\vert_0 -\frac \ii{12}
\bar\psi_{b}{}^{\dot\alpha}\,   \bar{\cal D}_{\dot\alpha}G^a\vert_0    \nn \\
&  - \frac \ii{64}
(\bar\psi_b\tilde\sigma^a\sigma_c)_{\dot{\alpha}}\,   \bar{\cal D}{}^{\dot\alpha}G^c\vert_0- \frac \ii{64}
(\psi_b\sigma^a\tilde\sigma_c)^{\alpha}\,   {\cal D}_{\alpha}G^c\vert_0
\Big]  \; , \nn\qquad\\
\label{E5}
E^a_{(5)}=&
\frac \ii{6} \, \theta\theta\, \bar{\theta} \bar{\theta} \, \Big[ (\mathcal T^{ab}+\frac \ii2 \varepsilon^{abcd}\mathcal T_{cd}) \sigma_b\d\bar \theta - \text{c.c.}
\Big]\;  .
\end{align}
\end{subequations}
where $\mathcal D_{\dot\alpha}$ is the spinorial covariant derivative defined in Appendix B and the expressions for the derivatives of $R$ and $G$, like $\mathcal DG|_0$, ${\mathcal D}\mathcal D G|_0$, ${\mathcal D}\mathcal D R|_0$ etc., in terms of space-time fields are given in Appendix \ref{DDR=fields}.

It is  important to note that the auxiliary fields enter $E^a$ starting from the second order in $\vartheta$. More precisely, the contribution of the auxiliary fields  to $E^a_{(2)}$ is linear, while in higher order terms of  $E^a$ they appear at most quadratically.
Furthermore, there are only linear terms
in space-time derivatives of the auxiliary fields and they appear only at the quartic order. This means that in the 3-brane action \eqref{3baction} the auxiliary fields appear only linearly or quadratically and without derivatives (modulo integration by parts). As such, when the 3-brane action is coupled to the supergravity action $S_{\rm SG}$ in \eqref{adssugra+3b}, one can still explicitly solve the equations of motion of the auxiliary fields \eqref{eomaux1} and \eqref{eomauxG1} modified by the presence of the goldstino fields.

\subsection{Action to the second order in goldstino}
Using the results of the previous subsection, one can straightforwardly write down the complete explicit expression of the 3-brane action $S_{\rm VA}$. However, in order to capture its main physical implications, we can just focus on the terms of this action which are linear and quadratic in the goldstino fields:
\be\label{3baction2}
\begin{aligned}
S_{\rm VA}=&f^2\int \d^4x\, e\, \Big[1+2\ii(\theta\sigma^a\bar{\psi}_a -
\psi_a\sigma^a\bar{\theta}) +
\ii(\theta\sigma^a\hat{\nabla}_a\bar{\theta} -
\hat{\nabla}_a\theta\sigma^a\bar{\theta})-\frac 18
G_a\theta\sigma^a\bar{\theta} \\
& +\frac 12 (\theta\theta \bar{R} + \bar{\theta}\bar{\theta} R) +
\bar{\theta} \bar{\theta} \psi_a\sigma^{ab}\psi_b + \theta\theta
\bar{\psi}_a\tilde{\sigma}^{ab}\bar{\psi}_b - 8\ii
 \varepsilon^{abcd}\theta\sigma^a\bar{\theta}\, {\psi}_b\sigma_c
\bar{\psi}_d \Big]+\ldots
\end{aligned}
\ee
 Upon substituting \eqref{3baction2} into  \eqref{adssugra+3b} and varying the latter with respect to $\bar R$ and $G^a$  we find the following solutions of the auxiliary field equations \eqref{eomaux1} and \eqref{eomauxG1}:
\be\label{solveRG}
R=-4m-\frac{8\kappa^2f^2}3\theta^2+\ldots,\qquad G^a={4\kappa^2f^2\over 3} \theta\sigma^a\bar{\theta}+\ldots.
\ee
Substituting these solutions back into the action \eqref{adssugra+3b} we get
\be\label{sgva}
\begin{aligned}
S=&\frac 1{2\kappa^2}\int\d^4 x\, e\Big[\mathcal R(\hat\omega) - 4 e^{-1}\varepsilon^{mnkl} (\hat\nabla_n\psi_{k}\sigma_{l} \bar{\psi}_m +{\psi}_m \sigma_n\hat\nabla_k\bar\psi_l)\ -4m(\bar{\psi}{}^a\sigma_{ab}\bar{\psi}{}^b+{\psi}{}^a\sigma_{ab}{\psi}{}^b)\Big]\\
&+\int \d^4 x\,e\Big\{\Big(f^2-\frac{3m^2}{\kappa^2}\Big) + f^2\Big[2\ii(\theta\sigma^a\bar{\psi}_a - \psi_a\sigma^a\bar{\theta}) +  \ii(\theta\sigma^a\hat\nabla_a\bar{\theta} - \hat\nabla_a\theta\sigma^a\bar{\theta}) - 2m(\theta^2+\bar\theta^2)\Big]\\
& +
f^2\left[\bar{\theta} \bar{\theta} \psi_a\sigma^{ab}\psi_b + \theta\theta
\bar{\psi}_a\tilde{\sigma}^{ab}\bar{\psi}_b - 8\ii
\varepsilon^{abcd}\theta\sigma_a\bar{\theta} \, {\psi}_b\sigma_c
\bar{\psi}_d\right]-\frac{3\kappa^2f^4}2\theta^2\bar\theta^2\Big\}+\ldots\,.
\end{aligned}
\ee
Modulo our conventions (see footnote \ref{resc}), (\ref{sgva}) coincides with the action of \cite{Bergshoeff:2015tra,Hasegawa:2015bza} truncated to the second order
in the goldstino.\footnote{In the Lagrangian \eqref{sgva} we also included the term $-\frac{3\kappa^2f^4}2\theta^2\bar\theta^2$ which is the only quartic $\vartheta$-term with constant coefficient (the same as in \cite{Bergshoeff:2015tra,Hasegawa:2015bza}), while the other quartic terms will contain supergravity fields.}

One can fix the local supersymmetry by imposing  the unitary gauge in which the goldstino vanishes:
\be\label{unitarygauge}
\theta(x)=0\,,
\ee
In this gauge,  the action \eqref{sgva} drastically simplifies:
\be\label{sgvaunitary}
\begin{aligned}
S=&\frac 1{2\kappa^2}\int e\Big[\mathcal R(\hat\omega) - 4 e^{-1}\varepsilon^{mnkl} (\hat\nabla_n\psi_{k}\sigma_{l} \bar{\psi}_m +{\psi}_m \sigma_n\hat\nabla_k\bar\psi_l)\\ &\quad~~~~~~~~~~ -4m(\bar{\psi}{}^a\sigma_{ab}\bar{\psi}{}^b+{\psi}{}^a\sigma_{ab}{\psi}{}^b)+\big(2\kappa^2 f^2-6 m^2\big)\Big]\,.
\end{aligned}
\ee
Notice that,  in the unitary gauge, the higher order contributions which we have omitted in (\ref{sgva}) vanish
and then (\ref{sgvaunitary}) provides the complete gauge-fixed action.  Hence, in this gauge the effect of the Volkov-Akulov 3-brane is reduced to a positive contribution to the value of the cosmological constant
\be
\Lambda=f^2-\frac{3m^2}{\kappa^2}
\ee
which can thus be positive and may give rise to a de Sitter vacuum as in \cite{Bergshoeff:2015tra,Hasegawa:2015bza}.

We expect that at higher orders in $\theta(x),\bar\theta(x)$ the form of our action will differ from that of \cite{Bergshoeff:2015tra,Hasegawa:2015bza} and similar constructions based on the nilpotent goldstino superfield in the same way as the original rigid-supersymmetry Volkov-Akulov action differs from its `constrained superfield' counterparts \cite{Ivanov:1977my,Ivanov:1978mx,Rocek:1978nb,Lindstrom:1979kq,Samuel:1982uh,Casalbuoni:1988xh,Komargodski:2009rz,Komargodski:2010rb}.
To find the precise correspondence between the two formulations one should generalize the non-linear field redefinitions relating different forms of the Volkov-Akulov Lagrangian found in \cite{Kuzenko:2010ef,Kuzenko:2011tj}.

\subsection{Emergence of constrained superfields}

Let us now demonstrate how a constraint on the chiral scalar superfield $R(z)=R(x,\theta,\bar\theta)$ entering the supergravity torsion and curvature arises in this formulation as a consequence of its equation of motion.
The expression \eqref{solveRG} for the complex scalar $R(x)$ to the second order in $\theta,\bar\theta$ is
\be\label{R=tt}
R(x)+4m=-\frac{8\kappa^2f^2}3\theta^2(x)\, (1+\ldots),
\ee
which implies that $R(x)$ is nilpotent, i.e.\ $(R(x)+4m)^2$=0. Since $R(x)$ is the lowest component of the chiral superfield $R(z)$, eq.\ \eqref{R=tt} may be regarded as  part of the solution to
 a nilpotency constraint involving  the entire superfield  $R(z)$ similar to that used in  \cite{Antoniadis:2014oya,Dudas:2015eha} to construct `nilpotent supergravity' models. This constrained was also derived in \cite{Kuzenko:2015yxa} as an equation of motion of supergravity coupled to the nilpotent chiral superfield.
As we will now show, this constraint is automatically solved by the superfield $R(z)$ obeying its equation of motion also in our formulation.

 Different ways to obtain superfield supergravity equations from superspace action by varying constrained supervielbeins were described in
\cite{Wess:1978bu,Buchbinder:1995uq,Siegel:1999ew,Bandos:2002bx} and references therein. With the use of the procedure of \cite{Bandos:2002bx} we get

\begin{eqnarray}\label{R=SfEq}
R(z)+4m =  \frac{16\kappa^2f^2}3\; {\cal J}(z)\; ,
\end{eqnarray}
where
\begin{eqnarray}\label{J=Sf}
{\cal J}(z) = (\bar{{\cal D}}\bar{{\cal D}} - R(z)) {\cal P}(z)\; , \qquad &
 {\cal P}(z) = \int \d^4\xi \frac{\det{\mathbb E}(z(\xi))}{{\rm Ber} \,E(z(\xi))} \; \delta^8(z-z(\xi))& \, ,   \qquad \end{eqnarray}
 and
 \begin{eqnarray}\label{delta=Sf} \delta^8(z) :=  {1\over 4} \, \theta^2  \, \bar\theta ^2\, \delta^4(x)  \qquad
\end{eqnarray}
is the superspace $\delta$ function obeying $\int \d^8 z\, \delta^8(z) h(z) =h(0)$. Notice that
$\bar{{\cal D}}\bar{{\cal D}} - R(z)$ is the chiral projector, \emph{i.e.}
 \begin{eqnarray}\label{delta=Sf1} \bar{{\cal D}}_{\dot{\alpha}}(\bar{{\cal D}}\bar{{\cal D}} - R(z)) {\cal P}(z)\equiv 0 \; , \qquad
\end{eqnarray}
Therefore, the right hand side of (\ref{R=SfEq}) is chiral, as it should be, because $R(z)$ is a chiral superfield.

Fixing the Wess-Zumino and the static gauge one finds that
\begin{eqnarray}\label{J=WZ}
{\cal J}(z) = - {1\over 2} \, (\theta - \theta(x))^2  \, \left( 1+ \ldots \right)\;    \quad \Rightarrow \quad {\cal J}(z)^2=0
\end{eqnarray}
Then, due to  equation (\ref{R=SfEq}) we get
\begin{eqnarray}\label{R-2=0}
(R(z)+4m)^2=0 \; .
\end{eqnarray}
The form of this constraint is the same as one obtained in \cite{Kuzenko:2015yxa} and  is similar to that used in \cite{Dudas:2015eha} for the construction of nilpotent supergravity
$$
(R(z)-\lambda)^2=0\,,
$$
However, the difference is that the parameter $\lambda$, which triggers supersymmetry breaking in the model of \cite{Dudas:2015eha}, is (\emph{a priori}) not related to the gravitino ``AdS mass" $m$.

\section{A comment on the Volkov-Soroka supergravity action}\label{VS}
In 1973 Volkov and Soroka \cite{Volkov:1973jd,Volkov:1974ai} coupled the Volkov-Akulov model to a simple supergravity multiplet consisting of the graviton and gravitino, and to a vector gauge field.
We will now briefly sketch their construction omitting the coupling to the vector fields (for a more detailed review see \cite{Volkov:1994te}) and show that in the unitary gauge the Volkov-Soroka action is the same as the one considered above.

The Volkov-Soroka model was based on a general approach to the construction of phenomenological Lagrangians with non-linearly realized symmetries \cite{Coleman:1969sm,Callan:1969sn,Volkov:1973vd} and consisted in gauging the super-Poincar\'e group by introducing corresponding gauge fields $e^{a}_{m}(x)$, $\psi^\alpha_m(x)$, $\bar\psi^{\dot\alpha}_m(x)$ and the Lorentz-algebra valued connection $\tilde\omega^{ab}_m(x)$ considered to be independent fields as in the first-order formalism to gravity and supergravity. The local supersymmetry transformations of $e^{a}_{m}(x)$, $\psi^\alpha_m(x)$ and $\bar\psi^{\dot\alpha}_m(x)$ which can be deciphered from \cite{Volkov:1973jd,Volkov:1974ai} have the same form as in \eqref{susy} with the auxiliary fields set to zero,
\be\label{susyvs}
\begin{aligned}
&\delta e^a_m=2\ii(\epsilon\sigma^a\bar\psi_m-\psi_m\sigma^a\bar\epsilon)\,,&\\
&\delta\psi_m=\tilde\nabla_m\epsilon,\qquad \delta\bar\psi_m=\tilde\nabla\bar\epsilon\,,&
\end{aligned}
\ee
but with the covariant derivative $\tilde\nabla=\d-\tilde \omega$ containing the independent connection $\tilde \omega_m(x)$. In the Volkov-Soroka construction $\tilde \omega_m(x)$ is invariant under supersymmetry transformations.

The Volkov-Soroka procedure of gauging the super-Poincar\'e group was essentially based on the idea to use the goldstino as a Stueckelberg-like field whose variation \hbox{$\delta\vartheta=-\epsilon(x)$} compensates the local supersymmetry transformations of the graviton and the gravitino together with a Stueckelberg-like field $X^a(x)$ to compensate local Poincar\'e translations in the tangent space. The latter can be gauge-fixed to zero thus reducing the action of the Poincar\'e group in the tangent space to local Lorentz rotations. Superinvariant one-forms constructed in this way (compare with \eqref{se2} and \eqref{Ea=WZ2})
\be
\begin{aligned}
\label{tildee}
&\tilde \psi=\psi+\tilde\nabla\theta,\qquad \bar{\tilde \psi}=\bar\psi+\tilde\nabla\bar\theta\,,&\\
&\tilde e^a(x)=e^a(x)+ 2\ii\theta\sigma^a\bar{\psi} - 2\ii\psi\sigma^a\bar{\theta}  +  \ii\theta\sigma^a\tilde\nabla \bar{\theta} -  \ii\tilde\nabla\theta\sigma^a\bar{\theta}\,&
\end{aligned}
\ee
were used to construct an invariant action
\be\label{VSL}
S_{{\rm VS}}=\frac {1}{2\kappa^2}\int\d^4 x\, \tilde e\Big[\mathcal R(\tilde\omega) - \frac{4c}{\tilde e}\varepsilon^{mnkl} (\tilde\nabla_n\tilde\psi_{k}\sigma_{l} \bar{\tilde\psi}_m +{\tilde\psi}_m \sigma_n\tilde\nabla_k\bar{\tilde\psi}_l) -4m(\bar{\tilde\psi}{}^a\sigma_{ab}\bar{\tilde\psi}{}^b+{\tilde\psi}{}^a\sigma_{ab}{\tilde\psi}{}^b)+\lambda\Big].\\
\ee
Note that in this action the coefficients $c$, $m$ and $\lambda$ are arbitrary.

Now, we can use the original supersymmetry transformations to put the goldstino field in this Lagrangian to zero, absorb the constant $c$ in the re-scaled gravitino field $(\psi \rightarrow c^{-\frac 12}\psi)$, redefine $\lambda=2\kappa^2(f^2-\frac{3m^2}{\kappa^2})$ and finally substitute the solution of the $\tilde\omega$ field equation back in to the action \eqref{VSL}{\footnote{It is important that in this construction $\tilde\omega$ is \emph{a priori} an independent field. Otherwise the re-scaling of the gravitino would not be possible.}.  The result is the gauge-fixed action \eqref{sgvaunitary}.

Back to 1973, when constructing their supergravity action Volkov and Soroka were, probably, too much concentrated on the local supersymmetry breaking associated with the shift of the goldstino and the corresponding super-Higgs effect, so they did not pose the question whether (for a suitable choice of the parameters) their action can still be supersymmetric even when the goldstino field is gauge fixed to zero.

\section{Discussion and outlook}
We have derived the minimal model describing the spontaneous breaking of pure $N=1$, $D=4$ supergravity induced by a space-filling 3-brane which carries the Volkov-Akulov goldstino on its worldvolume and provides a tunable constant  contribution to the cosmological constant which can be made positive. To the quadratic order in goldstino, the action coincides with previous constructions using constrained goldstino superfields \cite{Bergshoeff:2015tra,Hasegawa:2015bza}. We have also shown that  this action is equivalent to the Volkov-Soroka model \cite{Volkov:1973jd,Volkov:1974ai} in the unitary gauge in which goldstino vanishes.

This model is very naturally formulated in superspace and is based on the fact that the  Volkov-Akulov goldstini can be associated with the fluctuations of branes along Grassmann-odd directions in superspace. In this framework the model can be directly generalized to describe manifestly supersymmetric coupling of the Volkov-Akulov goldstino to more complicated supergravity-matter systems. For instance, one can consider a supergravity interacting with chiral scalar and vector supermultiplets described, respectively, by superfields $\Phi(z)$ and $V(z)$. These fields can couple to the 3-brane via their pull-back to the brane worldvolume. A straightforward generalization of action (\ref{action}) for such a system is
\bea\label{mattercoupling}
S&=&\frac 3{4\kappa^2} \int {\rm d}^8z \,{\rm Ber} \,E\,e^{-\frac{\kappa^2}3K(\bar\Phi\,e^{V},\,e^{V}\Phi)}+ \frac m{2\kappa^2}\Big(\int {\rm d}^6\zeta_L \,{\mathcal E}\,\big[\,W(\Phi)+{\rm tr}\,  g(\Phi)\,\mathbb W^\alpha \mathbb W_\alpha\big]+ {\rm c.c.}\Big) \nn\\
&&+{f^2} \int {\rm d}^4\xi \,\det\big[\mathbb{E}(z(\xi))\big]\, \mathcal F_{\Phi,\bar\Phi,V}(\xi)
\; ,
\eea
where $\mathcal F_{\Phi,\bar\Phi,V}(\xi)\equiv\mathcal F[\Phi(z(\xi)),\bar\Phi(z(\xi)),V(z(\xi))]$ is a real gauge-invariant function of the pull-backs of bulk superfields and their derivatives, while the first two terms describe the standard coupling of supergravity to  matter fields with a K\"ahler potential $K$, a superpotential $W$ and a complexified gauge coupling $g(\Phi)$. When reduced to the component field action, eq. \eqref{mattercoupling} should produce an alternative description of matter coupled supergravity with constrained superfields considered \emph{e.g.} in \cite{Hasegawa:2015bza,Kuzenko:2015yxa,Kallosh:2015sea,Kallosh:2015tea,Dall'Agata:2015zla,Schillo:2015}.

This framework allows to construct quite general Lagrangians which should provide, in a more direct way, low-energy effective field theories for string compactifications with anti-brane induced supersymmetry breaking. For instance, in KKLT-like scenarios, one may start from a generalization of (\ref{mattercoupling}) including additional worldvolume fields $\varphi(\xi)$, $A_i(\xi)$ etc. and match the brane term in (\ref{mattercoupling}) with what one obtains by dimensional reduction of the action of a probe anti-D3-brane sitting on an orientifold. In particular, the low-energy 3-brane tension $f^2$ would be proportional to $e^{4A_0} T_{\overline{\text{D3}}}$, where $T_{\overline{\text{D3}}}$ is the microscopic anti-D3-brane  tension and $e^{A_0}$ is the warp-factor of the ten-dimensional solution at the position of the probe anti-D3-brane.  As emphasized in \cite{Kachru:2003aw},  in order to get a sufficiently small effective 3-brane tension and make the low-energy effective field theory trustable, $e^{A_0}$  must be strongly suppressed compared to the average value of the warping along the compactification space.

Notice that such an approach to match the four-dimensional effective theory with the KKLT-like microscopic configuration would be based on the assumption that the action of a probe anti-brane captures the relevant physical information to derive the appropriate four-dimensional effective theory describing the fully back-reacted configuration.
In the recent years, starting with \cite{Bena:2009xk}, the validity of such constructions has been  under discussion, see \emph{e.g.}\ \cite{Cohen-Maldonado:2015ssa,Polchinski:2015bea} for a recent review of this problem and references therein for more details. The supersymmetric models that can be constructed with the use of the back-reacting Volkov-Akulov brane considered in this paper may be useful for tackling this issue from the perspective of the four-dimensional low-energy effective theory.

\subsection*{Acknowledgements}
The authors are grateful to Vladimir Akulov, Iosif Bena, Gianguido Dall'Agata, Fotis Farakos, Sergei Kuzenko, Kurt Lechner, Ichiro Oda and Paolo Pasti for useful discussions and comments. The work  of  I.~B.~has been supported in part by the Spanish MINECO grant FPA2012-35043-C02-01  partially financed  with FEDER funds,  by the Basque Government research group grant ITT559-10 and the Basque Country University program UFI 11/55. The work of L.~M.\ was partially supported by the Padua University
Project CPDA144437. The work of D.~S.~was partially supported by the Russian Science Foundation grant 14-42-00047 in association with Lebedev Physical Institute.

\appendix
\setcounter{equation}0
\def\theequation{\ref{notation}.\arabic{equation}}

\section{Notation and conventions}\label{notation}

We use the two-component Weyl-spinor formalism with the relativistic Pauli matrices $\sigma^a_{\beta\dot{\alpha}}= \epsilon_{\beta\alpha}
\epsilon_{\dot{\alpha}\dot{\beta}}
 \tilde{\sigma}{}^{a\dot{\beta}\alpha}$
obeying
 \be\label{sasb=}
 \sigma^a\tilde{\sigma}{}^b =\eta^{ab} +{\ii\over 2}\varepsilon^{abcd}\sigma_c\tilde{\sigma}_d\; ,\qquad
 \tilde{\sigma}{}^a{\sigma}^b =\eta^{ab} -{\ii\over 2}\varepsilon^{abcd}\tilde{\sigma}_c{\sigma}_d\; ,\qquad
 \qquad
\ee
where $\eta^{ab}={\rm diag} (1,-1,-1,-1)$ is the Minkowski metric and $\varepsilon^{abcd}$ is the totally
antisymmetric tensor with $\varepsilon^{0123}=1=-\varepsilon_{0123}$.
We rise and lower the spinorial indices $\theta_\alpha =\varepsilon_{\alpha\beta}\theta^\beta$ and
$\theta^\alpha =\varepsilon^{\alpha\beta}\theta_\beta$ with the use of
$\varepsilon^{\alpha\beta}=-\varepsilon^{\beta\alpha}= -\varepsilon_{\alpha\beta}$ obeying
$\varepsilon_{\alpha\beta}\varepsilon^{\beta\gamma}=\delta_\alpha^\gamma$.

The indices are contracted as follows
\be
\begin{aligned}
\label{thth=}
&\epsilon\sigma^a\bar{\psi} := \epsilon^\alpha  \sigma^a_{\alpha\dot\beta}\bar{\psi}
^{\dot\beta}, \qquad\bar{\psi}\tilde{\sigma}^{a} \epsilon:=\bar{\psi}_{\dot\beta}\tilde{\sigma}^{a\dot{\beta} \alpha } \epsilon_\alpha ,&\\
&\theta\theta=\theta^\alpha\theta_\alpha =  \varepsilon_{\alpha\beta} \theta^\alpha\theta^{\beta}\; \qquad
\bar{\theta}\bar{\theta}=\bar{\theta}_{\dot{\alpha}}\bar{\theta}^{\dot{\alpha}} = - \varepsilon_{\dot{\alpha}\dot{\beta}} \bar{\theta}^{\dot{\alpha}}\bar{\theta}^{\dot\beta}\;  ,\qquad \;
\theta\theta=(\bar{\theta}\bar{\theta})^*\;.&
\quad
\end{aligned}
\ee
The antisymmetrized products of the Pauli matrices
 $$\sigma^{ab}{}_{\beta} {}^\alpha=\sigma^{[a}\tilde{\sigma}^{b]}{}_{\beta} {}^\alpha:=
{1\over 2}(\sigma^{a}\tilde{\sigma}^{b}- \sigma^{b}\tilde{\sigma}^{a})_{\beta} {}^\alpha, \qquad
\tilde{\sigma}{}^{ab}{}^{\dot\alpha}{}_{\dot\beta}=\tilde{\sigma}^{[a}{\sigma}^{b]}{}^{\dot\alpha}{}_{\dot\beta} $$
are, respectively, imaginary-self-dual and imaginary-anti-self-dual,
$$\sigma^{ab}={\ii\over 2}\varepsilon^{abcd}\sigma_{cd},\qquad  \tilde{\sigma}{}^{ab} =-{\ii\over
2}\varepsilon^{abcd}\tilde{\sigma}_{cd}.$$
Some other useful properties are:
\begin{subequations}
\begin{align}
\label{s2s+ss2=esA}
\sigma_{ab}\sigma_c+\sigma_c\tilde{\sigma}{}_{ab}= -2\ii\varepsilon_{abcd} \sigma^d \; , \qquad \tilde{\sigma}{}_{ab}\tilde{\sigma}_c+\tilde{\sigma}_c{\sigma}{}_{ab}= 2\ii\varepsilon_{abcd} \tilde{\sigma}^d  \; .    \quad \\
\label{s2s-ss2=sA}
\sigma_{ab}\sigma_c-\sigma_c\tilde{\sigma}{}_{ab} = 4{\sigma}_{[a}\eta_{b]c} \; , \qquad \tilde{\sigma}{}_{ab}\tilde{\sigma}_c-\tilde{\sigma}_c{\sigma}{}_{ab}= 4\tilde{\sigma}_{[a}\eta_{b]c}  \; ,    \quad
 \\
\label{sss-sss=esA}
\sigma_{b}\tilde{\sigma}_a\sigma_c- \sigma_c\tilde{\sigma}_a\sigma_b= 2\ii\varepsilon_{abcd} \sigma^d \; , \qquad \tilde{\sigma}_{b}{\sigma}_a\tilde{\sigma}_c- \tilde{\sigma}_c{\sigma}_a\tilde{\sigma}_b= -2\ii\varepsilon_{abcd} \tilde{\sigma}^d \; .   \quad
  \end{align}
  \end{subequations}

\setcounter{equation}0
\def\theequation{\ref{constraints}.\arabic{equation}}

\section{$N=1$, $D=4$ supergravity constraints}\label{constraints}
The constraints on the torsion (the differentials act from the right)
\be\label{torsion}
\mathcal DE^A=\d E^A-E^B\wedge \Omega_B{}^A=T^A
\ee
are
\begin{subequations}\label{solvedtorsion}
\begin{align}
\label{4WTa=} & T^a
=- 2\ii\sigma^a_{\alpha\dot{\alpha}} E^\alpha \wedge \bar{E}^{\dot{\alpha}} -{1\over 8} E^b \wedge E^c
\varepsilon^a{}_{bcd} G^d \; ,  \hspace{4.0cm} \\ \label{4WTal=} & T^{\alpha}  = {\ii\over 8} E^c \wedge E^{\beta}
(\sigma_c\tilde{\sigma}_d)_{\beta} {}^{\alpha} G^d   -{\ii\over 8} E^c
\wedge \bar{E}^{\dot{\beta}} \varepsilon^{\alpha\beta}\sigma_{c\beta\dot{\beta}}R +
 {1\over 2} E^c \wedge E^b \; T_{bc}{}^{\alpha} \; , \\
\label{4WTdA=} & T^{\dot{\alpha}}  = {\ii\over 8} E^c \wedge E^{\beta} \varepsilon^{\dot{\alpha}\dot{\beta}}
\sigma_{c\beta\dot{\beta}} \bar{R}  -{\ii\over 8} E^c \wedge
\bar{E}^{\dot{\beta}} (\tilde{\sigma}_d\sigma_c)^{\dot{\alpha}}{}_{\dot{\beta}} \, G^d +
 {1\over 2} E^c \wedge E^b \; T_{bc}{}^{\dot{\alpha}}\; .
\end{align}
\end{subequations}
In these expressions  $G_a(z)$ and $R(z)=(\bar{R}(z))^*$ are so-called main off-shell superfields of $N=1$ $D=4$ supergravity, which obey a number of relations including
\begin{subequations}
\begin{align}
\label{chR} & {\cal D}_\alpha \bar{R}=0\;
, \qquad \bar{{\cal D}}_{\dot{\alpha}} {R}=0\; ,
 \\
\label{DG=DR} & \bar{{\cal
D}}^{\dot{\alpha}}G_a\sigma^a_{{\alpha}\dot{\alpha}}= -{\cal D}_{\alpha} R \; , \qquad {{\cal
D}}^{{\alpha}}G_a\sigma^a_{{\alpha}\dot{\alpha}}= -\bar{{\cal D}}_{\dot{\alpha}} \bar{R} \; . \qquad
\end{align}
\end{subequations}
If we collectively denote the fermionic coordinates by
$\vartheta^{\hat\mu}=(\theta^\mu,\bar\theta^{\dot\mu})$, the lowest $\vartheta=0$ components  of these superfields are the auxiliary fields of minimal supergravity multiplet, which we will denote by the same symbol,
$G_a(x)\equiv  G_a(z)\vert_{\vartheta=0}$, $R(x)\equiv R(z)\vert_{\vartheta=0}$.

Notice that our covariant derivatives are defined from the following decomposition of the covariant differential
\begin{eqnarray} \label{D=ED}
{\cal D}=E^a{\cal D}_{a} +E^{\alpha} {\cal D}_{\alpha}  +
\bar{E}{}^{\dot{\alpha}} \bar{{\cal D}}_{\dot{\alpha}} \; . \qquad \end{eqnarray}
Since complex conjugation interchange the order of fermionic entities, this implies
 \begin{eqnarray} \label{bD*-D}
 (\bar{{\cal D}}_{\dot{\alpha}})^*=- {\cal D}_{\alpha}\; . \qquad \end{eqnarray}

The complete set of the main superfields also includes  the symmetric spin-tensor  ${W}^{{\alpha}{\beta}{\gamma}}:= 4 \tilde{\sigma}{}^{ab(\alpha\beta} T_{ab}{}^{\gamma   )}$ and its complex conjugate
$\bar{W}{}^{\dot{\alpha} \dot{\beta}\dot{\gamma}}= - 4 \tilde{\sigma}{}^{ab(\dot{\alpha}\dot{\beta}} T_{ab}{}^{\dot{\gamma} )}$ which are chiral
\begin{eqnarray}
\label{chW}  \bar{{\cal D}}_{\dot{\alpha}} W^{\alpha\beta\gamma}= 0\; , \qquad {{\cal D}}_{{\alpha}}
\bar{W}^{\dot{\alpha}\dot{\beta}\dot{\gamma}}= 0\; , \end{eqnarray}
and obey the relations
\be \label{DW=DG}
{{\cal D}}_{{\gamma}}W^{{\alpha}{\beta}{\gamma}}= \bar{{\cal D}}_{\dot{\gamma}} {{\cal
D}}^{({\alpha}}G^{{\beta})\dot{\gamma}} \; , \qquad \bar{{\cal D}}_{\dot{\gamma}}
\bar{W}^{\dot{\alpha}\dot{\beta}\dot{\gamma}} = {{\cal D}}_{{\gamma}} \bar{{\cal D}}^{(\dot{\alpha}|}
G^{{\gamma}|\dot{\beta})} \; . \qquad \ee
The symmetric part of the non-vanishing Grassmann covariant derivative of $W_{\alpha\beta\gamma}$ produces the superfield generalization of the irreducible
(spin-tensor) components of the Weyl tensor
  \be
\label{Cffff=Rss}
 C_{\alpha\beta \gamma\delta} :=
C_{(\alpha\beta{\gamma}{\delta})} =  {1\over 16} \sigma_{ab (\alpha\beta}   \sigma^{cd}_{\gamma\delta)}
\mathcal R_{cd}{}^{ab}
  \ee
 and its c.c.,
\begin{eqnarray} \label{Weyl}
{{\cal D}}_{({\alpha}} W_{\beta{\gamma}{\delta})}=- 16 C_{\alpha\beta{\gamma}{\delta}} \; , \qquad
 {{\cal D}}_{(\dot{\alpha}} \bar{W}_{\dot{\beta}\dot{\gamma}\dot{\delta})} =-16 \bar{C}_{\dot{\alpha}\dot{\beta}\dot{\gamma}\dot{\delta}}\; .
\end{eqnarray}
The superfield curvature is defined in terms of the main superfields as follows
\be\label{scurv}
\mathcal R^{ab}=\d\Omega^{ab}-{\Omega}^{ac}\wedge {\Omega}_c{}^{b}=\frac 12 \mathcal R^{\alpha\beta}(\sigma^a\tilde\sigma^b)_{\alpha\beta}-\frac 12 \mathcal R^{\dot\alpha\dot\beta}(\tilde\sigma^a\sigma^b)_{\dot\alpha\dot\beta}
\ee
and
\be
\begin{aligned}
\label{spincurv}
\mathcal R^{\alpha\beta}=&\d{\Omega}^{\alpha\beta}-{\Omega}^{\alpha\gamma}\wedge  {\Omega}_{\gamma}{}^\beta=\frac 14 \mathcal R^{ab}(\sigma_a\tilde\sigma_b)^{\alpha\beta}\\
=&-\frac 12 E^\alpha\wedge E^\beta \bar R-\frac \ii8 E^c\wedge E^{(\alpha}\tilde\sigma_c^{\dot\gamma\beta)}\bar{\mathcal D}_{\dot\gamma}\bar R+\frac \ii8 E^c\wedge E^\gamma(\sigma_c\tilde\sigma_d)_\gamma{}^{(\beta}\mathcal D^{\alpha)}G^d\\
&-\frac \ii8 E^c\wedge \bar E^{\dot\beta}\sigma_{c\gamma\dot\beta}W^{\alpha\beta\gamma}+\frac 12 E^d\wedge E^c \mathcal R_{cd}{}^{\alpha\beta}\,.
\end{aligned}
\ee
The form of the curvature is obtained by solving the Bianchi identities
\be\label{Bianchi}
\mathcal DT^A+E^B\mathcal \wedge R_B{}^A=0.
\ee

Note that in the chosen form of the torsion constraint \eqref{4WTa=} containing non-zero $T_{bc}^c$ components, the lowest component of the superfield connection ${\Omega}_A{}^{B}$ is related to the conventional supergravity spin connection
\be\label{conomega}
\hat\omega^{ab}_m:=\omega^{ab}_m+2\ii(\psi^{[a}\sigma^{b]}\bar{\psi}_m+\psi_{m}\sigma^{[a}\bar{\psi}^{b]}+\psi^{[a}\sigma_{m}
   \bar{\psi}^{b]})
\ee
as follows
\be
\begin{aligned}
\label{wab0=omab+}
{\Omega}_{m }^{ab}\vert_0=& \hat\omega_{m }^{ab}-{1\over 8} e_{m\, c}\varepsilon^{abcd} G_d\vert_0
 \\ {\Omega}_{m }{}_\alpha{}^{\beta}\vert_0=& \hat\omega_{m\alpha}{}^{\beta}-{1\over 32} e_{m\, c}\varepsilon^{abcd}(\sigma_a\tilde\sigma_b){}_\alpha{}^{\beta} G_d\vert_0=\hat\omega_{m\alpha}{}^{\beta}(x)+{i\over 16} e_{m}^a\, (\sigma_{[a}\tilde\sigma_{b]}){}_\alpha{}^{\beta} G^b\vert_0
 \, \; .
 \end{aligned}
 \ee
So, e.g. for an arbitrary spinor superfield $W^\alpha$
\be
\begin{aligned}
\label{Dnabla}
 \mathcal {\cal D}_{m }W^\alpha\vert_0 &=\hat\nabla_{m }W^\alpha\vert_0 +{1\over 32} e_{m\, c}\varepsilon^{abcd}(W\sigma_a\tilde\sigma_b) G_d\vert_0 =\\
 &=
\hat\nabla_{m }W^\alpha\vert_0 -{\ii\over 16} e_{m}^a(W\sigma_{[a}\tilde\sigma_{b]}) G^b\vert_0
 \,,
 \end{aligned}
 \ee
where
\be\label{nabla}
\hat\nabla W^{\alpha}\vert_0 =\d W^{\alpha}(x,0) -W^\beta{}(x,0) \hat\omega{}_{\beta}{}^{\alpha}
\ee
is a conventional covariant derivative acting on the component supergravity fields.

\section{Lowest components of superfields $R$, $\bar R$ and $G^a$ in terms of space-time fields}\label{DDR=fields}
\setcounter{equation}0
\def\theequation{\ref{DDR=fields}.\arabic{equation}}

One of the characteristic properties of the Wess-Zumino gauge is that, with the help of the supergravity constraints and their Bianchi identities, higher components of superfields can be written as lowest components of the fermionic covariant derivatives of these superfields. In particular, the first terms in the $\vartheta$ decomposition of the  superfields $R$, $\bar R$ and $G^a$ , which we need to solve equations \eqref{dea}-\eqref{debalpha} for $E^a$, are
\begin{subequations}
\begin{align}
\label{R1=}
R_{(0)}&=R(x)\; , \qquad R_{(1)}=\theta^\alpha  (\mathcal D_\alpha R)\vert_0=-\theta^\alpha \sigma^a_{\alpha\dot\beta}(\bar{\mathcal D}^{\dot\beta}G_a)\vert_0\, \; ,  \\ \label{R2=}
R_{(2)}&= -\frac 14 \theta\theta \; {\cal D}^\alpha {\cal D}_\alpha R\vert_0 + \frac 12 \theta^{\alpha}\bar{\theta}{}^{\dot\alpha} \bar{\cal D}_{\dot\alpha}  {\cal D}_\alpha R\vert_0\; ,  \\
\label{G2=}
G^{a}_{(0)}&=G^a(x)\; , \qquad G^a_{(1)}= \theta^\alpha {\cal D}_\alpha G^a\vert_0 +  \bar\theta{}^{\dot\alpha}  \bar{\cal D}_{\dot\alpha}  G^a\vert_0 \; , \\
G^a_{(2)}&= -\frac 14 \theta\theta \; {\cal D}^\alpha {\cal D}_\alpha G^a\vert_0 -\frac 14 \bar\theta\bar\theta\;  \bar{\cal D}_{\dot\alpha}  \bar{\cal D}{}^{\dot\alpha}  G^a\vert_0 +\frac 12 \theta^{\alpha}\bar{\theta}{}^{\dot\alpha} \; [\bar{\cal D}_{\dot\alpha}  , {\cal D}_\alpha ] G^a\vert_0\; ,\\
\label{W1=}
W^{\alpha\beta\gamma}\vert_0& =  4 {\sigma}{}^{ab({\alpha}{\beta}} T_{ab}{}^{{\gamma} )}\vert_0 \; ,  \qquad
W^{\alpha\beta\gamma}_{(1)}= \theta^\delta\;  {\cal D}_\delta  W^{\alpha\beta\gamma}\vert_0\; .
\end{align}
\end{subequations}
These expressions still cannot be used straightforwardly. First of all one should decompose covariant derivatives of the superfields on irreducible parts and identify them.

The fermionic derivatives of the main off-shell superfields are expressed by
\begin{eqnarray}\label{DfGa0=Psi}
\bar{{\cal D}}_{\dot\alpha} G^a\vert_0 = {4\ii} \big[  {{\cal T}}_{\dot{\alpha}}^a - {1\over 3}
({{\cal T}}^b\tilde{\sigma}^a{\sigma}_b )_{\dot{\alpha}}\big] \, , \qquad
\label{DfR0=Psi}
{} {\cal D}_\alpha R\vert_0 =  {4\ii\over 3} \sigma_{a\alpha\dot\alpha} {{\cal T}}^{\dot{\alpha}a}  \;
\end{eqnarray}
through the superfield generalization of the fermionic equation of supergravity  (see (\ref{RSEq:=}))
\begin{eqnarray}\label{PPsiada:=}
{{\cal T}}^a_{\dot{\alpha}}(x) =\varepsilon^{abcd} {\cal T}_{bc}{}^\alpha (x) \sigma_{d\alpha\dot{\alpha}}\;
\end{eqnarray}
where recall that
\begin{eqnarray}\label{Tabal:=1}
\mathcal T_{ab}{}^\alpha(x):=T_{ab}{}^\alpha\vert_0= \psi_{ab}^\alpha(x) -{1\over 8} (\psi_{[a}\sigma_{b]}\tilde\sigma^{c})^{\alpha}G_c(x) -
{\ii\over 8} \psi_{[a}{}^{{\alpha}}G_{b]}(x) - {\ii\over 4}
(\bar\psi_{[a}\tilde\sigma_{b]})_{\dot{\alpha}}R(x)
\end{eqnarray}
 is the lowest component of the fermionic superspace  torsion  and
\be\label{psiab1}
\psi_{ab}^\alpha(x)=2e_a^me_b^n\hat\nabla_{[m}\psi^\alpha_{n]}
\ee
The second term in (\ref{R2=}) is expressed through the bosonic derivative of the lowest component of the chiral superfield $R$ as
 \bea\label{bDDR=}
  \bar{\cal D}_{\dot\alpha}  {\cal D}_\alpha R\vert_0= \{ \bar{\cal D}_{\dot\alpha},  {\cal D}_\alpha \} R\vert_0=  2\ii {\cal D}_{\alpha\dot\alpha} R\vert_0= - 2\ii\sigma^a_{\alpha\dot\alpha} (e_a^m\partial_m R- \psi_a^\beta {\cal D}_\beta R\vert_0)\; .
 \eea
Furthermore,
using the consequences of the supergravity constraints we obtain, in particular, that
\be\label{Walbega0=1}
W^{\alpha\beta\gamma}\vert_0 =  4 {\sigma}{}^{ab({\alpha}{\beta}} \mathcal T_{ab}{}^{{\gamma} )}=  4  {\sigma}{}^{ab({\alpha}{\beta}} \psi_{ab}{}^{{\gamma} )}- \frac {3\ii}{2} {\sigma}{}^{ab({\alpha}{\beta}} \psi^{\gamma )}_{a}G_b
 \; .
\ee
and
 \be
\label{DfW=}
{\cal D}_\alpha W^{\beta\gamma\delta}= -16  C_{\alpha}{}^{\beta \gamma\delta}-   {3\ii\over 2}\delta_{\alpha}{}^{(\beta}
\sigma_{ab}^{\gamma\delta ) }{\cal D}^{a}G^b\;,  \qquad
\ee
where the lowest component of (\ref{DfW=}) is expressed in terms of the irreducible part of the Weyl tensor superfield (\ref{Cffff=Rss})
\be
\begin{aligned}
\label{Cffff0=2}
 C_{\alpha\beta \gamma\delta} \vert_0=&  {1\over 16} \sigma_{ab (\alpha\beta}   \sigma^{cd}_{\gamma\delta)}    e_c^m e_d^n  \mathcal R_{mn}{}^{ab} +    {\ii\over 8}(\sigma^c\bar{\psi}_c)_{(\alpha} W_{\beta \gamma\delta )}\vert_0   - {1\over 4}\bar{R}  \psi _{ a(\alpha}\sigma^{ab}_{\beta \gamma}\psi_{\delta)b}  \\ &- {\ii\over 16}\psi _{ a(\alpha}(\sigma^{a}\tilde{\sigma}{}^b)_{\beta \gamma}
  {\cal D}_{\delta)} G_{b}\vert_0+ {1\over 8} \psi _{ a(\alpha} \sigma^{ab}_{\beta\gamma} (\sigma^c\bar{\psi}_b)_{\delta)} G_{c}\vert_0 \;  .  \qquad
  \end{aligned}
  \ee
and
\be\label{DaG}
{\cal D}_{[a}G_{b]}\vert_0 = e_{[a}{}^m \hat\nabla_m G_{b]} \vert_0 - \psi_{[a}{}^\alpha \mathcal D_\alpha G_{b]} \vert_0  -  \bar{\psi}_{[a}{}^{\dot\alpha} \mathcal D_{\dot\alpha}  G_{b]} \vert_0 \; .
\ee
To obtain a more explicit expression in terms of space-time fields, $ {\cal D}_{\alpha} G_{a}\vert_0$ and
$ W_{\beta \gamma\delta}\vert_0$ should be specified with the use  of (\ref{DfGa0=Psi}) and (\ref{Walbega0=1}).

Some expressions for the second Grassmann derivatives of $G_a$ and $R$ are also useful.
Using (\ref{DG=DR}) and (\ref{chR}), after some algebra with covariant derivatives we find
\be
\label{sbDDG=}
{\cal D}^{\alpha}{\cal D}_{\alpha}G_a =4\ii {\cal D}_a\bar{R} + \frac 32 \bar{R}  G_a\; , \qquad  \bar{{\cal D}}_{\dot\alpha}\bar{{\cal D}}^{\dot\alpha} G_a =- 4\ii {\cal D}_a{R}+ \frac 32 {R}  G_a \; , \qquad
\ee
so that ${\cal D}^{\alpha}{\cal D}_{\alpha}G_a \vert_0$ and $\bar{{\cal D}}_{\dot\alpha}\bar{{\cal D}}^{\dot\alpha} G_a \vert_0$ can be calculated using (\ref{bDDR=}). Next, using
\be
\label{sbDDG=1}
\bar{{\cal D}}_{(\dot{\alpha}|}{\cal D}_{({\alpha}}G_{\beta )|\dot{\beta})}= \tilde{\sigma}^{cd}{}_{\dot{\alpha}\dot{\beta}} {\sigma}_{ab\, {\alpha}{\beta}} \hat R_{cd}{}^{ab}\;   \qquad
  \ee
 and equation (\ref{DG=DR}) we find that
\be
\label{bDDG=}
\bar{{\cal D}}_{\dot{\alpha}}{\cal D}_{{\alpha}}G_{\beta \dot{\beta}}= \tilde{\sigma}^{cd}{}_{\dot{\alpha}\dot{\beta}} {\sigma}_{ab\, {\alpha}{\beta}} R_{cd}{}^{ab} - i \epsilon_{\dot{\alpha}\dot{\beta}} \sigma^{ab}_{{\alpha}\beta}{}^{\dot{\gamma}}{\cal D}_{[a}G_{b]}+ {1\over 4}\epsilon_{{\alpha}{\beta}}\epsilon_{\dot{\alpha}\dot{\beta}}\bar{{\cal D}}\bar{{\cal D}}\bar{R}\;  .  \qquad
  \ee
  To obtain the final expression for  the lowest component $\bar{{\cal D}}_{\dot{\alpha}}{\cal D}_{{\alpha}}G_{\beta \dot{\beta}}\vert_0$ in terms of spacetime fields we have to take into account that
  \begin{eqnarray}
\label{Rabcd=R0+}
 R_{mn}{}^{ab}\vert_0 &= \mathcal R_{mn}{}^{ab} +  {1\over 4}\varepsilon^{abcd}e_{c[m}\hat\nabla_{n]}G_d- {\ii\over 2}\varepsilon^{abcd}\psi_{[m |}\sigma_c \bar{\psi}_{|n ]} G_d+  {3\over 32}e_m^{[a} e_n^b G^{c]} G_c. & \quad
  \end{eqnarray}
Finally,
\be\label{DDR0=-R0+}
\begin{aligned}
{} {\cal D}^\alpha {\cal D}_\alpha R\vert_0 =& - \frac{16\kappa^2}{ 3e} L_{\rm SG} +  3 R\bar{R} - 2\ii e_a^m \hat\nabla_m G^a + 8R \bar{\psi}_a\bar{\psi}^a- 4G_a{\psi}^b{\sigma}_b\bar{\psi}^a  \\ & + 8 \varepsilon^{abcd} ({\psi}_{bc}{} \sigma_{d} \bar{\psi}_a - 8 {\psi}_a\sigma_d\bar\psi_{bc})
- {16\over 3}\varepsilon^{abcd} {\psi}_{bc}\sigma_d\tilde{\sigma}_a{\sigma}_f \bar{\psi}^{f}\; , \qquad
\end{aligned}
\ee
where
$ L_{\rm SG}$ is the component field Lagrangian for minimal $N=1$, $D=4$ supergravity with zero cosmological constant
\be\label{LminSG}
{}  \frac{2\kappa^2}e L_{\rm SG} =  \mathcal R(\hat\omega) - 4 e^{-1}\varepsilon^{mnkl} (\hat\nabla_n\psi_{k}\sigma_{l} \bar{\psi}_m +{\psi}_m \sigma_n\hat\nabla_k\bar\psi_l)+ {3\over 8}  R\bar{R}   + {3\over 32}  G_aG^a .
\ee

\section{Supersymmetry transformations preserving the Wess-Zumino gauge}\label{WZg}
\setcounter{equation}0
\def\theequation{\ref{WZg}.\arabic{equation}}

The parameters $\varepsilon^M(x,\theta,\bar\theta) =z^M- z^{\prime M}$  and $L^{ab}(z)$ of the infinitesimal superdiffeomorphism and local Lorentz transformations $E^{\prime A}(z^{\prime})=E^A(z)- E^B(z)L_B{}^A(z)$  which preserve the Wess-Zumino gauge (\ref{WZgauge}) should obey the following system of equations
\be\label{WZgPresEq}
\begin{aligned}
\vartheta^{\hat\alpha} \partial_{\hat\alpha} (\iota_\epsilon E^A) + \vartheta^{\hat\alpha}\iota_\epsilon E^BT_{B\hat\alpha}{}^A +
 \vartheta^{\hat\beta}(L_{\hat\beta}{}^{\hat\alpha} + \iota_\epsilon \Omega_{\hat\beta}{}^{\hat\alpha}) \delta_{\hat\alpha}{}^A=0\; ,  \\
\vartheta^{\hat\rho} \partial_{\hat\rho}(L^{ab} + \iota_\epsilon w^{ab}) + \vartheta^{\hat\rho}\iota_\epsilon E^B\mathcal R_{B\hat\rho}{}^{ab}=0\;
 \; ,
\end{aligned}
\ee
where remember that $\vartheta^{\hat\mu}=(\theta^\mu,\bar\theta^{\dot\mu})$. Equivalently, as we did for finding the $\theta$-expansion of the supervielbein, one can perform the re-scaling $\vartheta\mapsto t\vartheta$ and write \eqref{WZgPresEq} as an equation for derivative in $t$,
\be\label{WZgPres=tEq}
\begin{aligned}
& \frac {\d}{\d t} \iota_\epsilon E^A(t) =- \vartheta^{\hat\alpha}\iota_\epsilon E^B(t)T_{B\hat\alpha}{}^A (t) -
 \vartheta^{\hat\beta}\tilde{L}_{\hat\beta}{}^{\hat{\alpha}}(t)\delta_{\hat{\alpha}}^A \; ,  \\
& \frac {\d}{\d t}\tilde{L}^{ab}(t) =- \vartheta^{\hat\rho}\iota_\epsilon E^B(t)\mathcal R_{B\hat\rho}{}^{ab}(t)\;
 \; ,
\end{aligned}
\ee
where $\tilde{L}^{ab}:= (L^{ab} + \iota_\epsilon \Omega^{ab})$.  It is easy to see  that these equations do not fix the lowest component of the superfield parameters, {\it i.e.}
\be\label{WZgPres=tEq0}
\iota_\epsilon E^a\vert_0= \epsilon^m (x) e_m^a(x)=: \epsilon^a(x)\; , \qquad \iota_\epsilon E^\alpha\vert_0= \epsilon^\alpha (x) + \epsilon^a(x)\psi_a^\alpha(x)\; , \qquad
\ee
and $ \tilde{L}^{\alpha\beta}\vert_0= {L}^{\alpha\beta}(x) + \epsilon^a(x)\Omega_a^{\alpha\beta}(x,0) $ remains free and correspond to the gauge symmetries of spacetime formulation of supergravity. In particular,
$\epsilon^\alpha (x)$ can be identified as the parameter of local spacetime supersymmetry.

To obtain the local supersymmetry transformation of the goldstino,
\be\label{dgold}
\delta \vartheta^{\hat{\mu}}(x)= -\epsilon^{\hat{\mu}} (x,\vartheta(x))- \delta_\epsilon x^m\partial_m\vartheta^{\hat\mu} = -\epsilon^{\hat{\mu}} (x,\vartheta(x))+ \epsilon^m (x,\vartheta(x))\partial_m\vartheta^{\hat\mu}
\ee
 we need to know the expressions for higher order terms in the decomposition of the
superfield parameter $\epsilon^{{\hat{\alpha}}} (x,{\vartheta})$ in the superspace Grassmann coordinates.
These can be found by solving eqs.  (\ref{WZgPres=tEq}) for $\iota_\epsilon E^A:=\epsilon^NE_N^A $ and then finding
$\epsilon^N(z)$ using the inverse of $E_N^A$. To this end we should know the explicit form of the $\vartheta$-decomposition of the supervielbein in the WZ gauge. Below we will present the expansion of $\epsilon^{\hat\alpha}(x,\theta, \bar\theta)$ up to the second order in $\theta,\bar\theta$.

With our choice of the supergravity constraints, eqs. (\ref{WZgPres=tEq})  take the form
\begin{subequations}
\begin{align}
\label{dEPa}
\frac \d{\d t}\iota_\epsilon E^a=&
2\ii(\theta\sigma^a\iota_\epsilon\bar E-\iota_\epsilon E\sigma^a\bar\theta)\,,\\
\label{dEPalpha}
\frac \d{\d t}\iota_\epsilon E^\alpha=&\frac \ii8 \iota_\epsilon E^c[(\theta\sigma_c\tilde \sigma_d)^\alpha G^d+\bar \theta_{\dot\beta}\tilde\sigma_c^{\dot\beta\alpha}  R] - \theta^{{\beta}}\tilde{L}_{{\beta}}{}^{{\alpha}}\,,\\
\label{dEPbalpha}
\frac \d{\d t}\iota_\epsilon E^{\dot\alpha}=&-\frac \ii8 \iota_\epsilon E^c[(\tilde \sigma_d\sigma_c\bar\theta)^{\dot\alpha} G^d+ \tilde\sigma^{c\dot\alpha\beta}\theta_{\beta}\bar R]- \bar{\theta}{}^{\dot{\beta}}\tilde{L}_{\dot{\beta}}{}^{\dot{\alpha}}\,.\\
\label{dEPw=4D}
\frac \d{\d t} \tilde{L}^{\alpha\beta}=& -  \iota_\epsilon  E^{(\alpha } \theta^{\beta
)}\bar{R} - \frac \ii8 \iota_\epsilon  E^c \left[
\tilde{\sigma}_c^{\dot{\gamma}(\alpha }\theta^{\beta )}\bar{{\cal
D}}_{\dot{\gamma}}\bar{R} - (\theta\sigma_c \tilde{\sigma}_d)^{(\alpha
} {\cal D}^{\beta )}G^d+ (\sigma_c\bar{\theta})_\gamma
W^{\alpha\beta\gamma} \right] .
\end{align}
\end{subequations}
To find the supersymmetry transformation with the independent parameter $\epsilon^{\hat\alpha}$ it is enough to solve \eqref{dEPa}--(\ref{dEPw=4D}) with $\epsilon^a(x)=0=\tilde{L}^{ab}(x)$
\be\label{bc-susy}
\iota_\epsilon E^a\vert_0(x)=0\; ,  \qquad
 \iota_\epsilon E^\alpha\vert_0(x)= \epsilon^\alpha (x) \; , \qquad \tilde{L}^{\alpha\beta}\vert_0= {L}^{\alpha\beta}(x)=0\; .
\ee
It is easy to find the solution up to the third  order in $\theta$
\be\label{ssp-susy-a}
\begin{aligned}
\iota_\epsilon E^a=& 2\ii(\theta\sigma^a\bar\epsilon-\epsilon\sigma^a\bar\theta)+
\frac \ii{12}\theta\theta\, \bar\theta_{\dot{\alpha}}\,\left[- (\tilde{\sigma}^a\epsilon)^
{\dot{\alpha}}\bar{R} + 4 \bar\epsilon{}^{\dot{\alpha}}G^a + 2(\tilde{\sigma}^{ab}\bar\epsilon)^{\dot{\alpha}}G_b\right] -\\
&-\frac \ii{12}
\bar\theta\bar\theta\, \theta^{\alpha}\,\left[({\sigma}^a\bar\epsilon)_{\alpha} R
+4 \epsilon_{\alpha} G^a + 2 ({\sigma}^{ab}\epsilon)_{\alpha} G_b\right] +  \calo( \vartheta^4)\, ,
\end{aligned}
\ee
which requires the knowledge of  $\iota_\epsilon E^\alpha$ to the second order in $\theta$
\be\label{ssp-susy-al}
\begin{aligned}
 \iota_\epsilon E^\alpha=& \epsilon^\alpha  -  \frac 1{4}\, \theta\theta\, \left(  (\bar\epsilon\tilde{\sigma}{}^a)^\alpha G_a+3\epsilon^\alpha \bar R \right) -  \frac 1{4}\,\epsilon^\alpha\, \bar\theta\bar\theta\, R \\ &+ \frac 1{4}\, \theta^\alpha\,  \bar\theta\bar\epsilon\, R+  \frac 1{4}\, \theta\epsilon\, (\bar\theta\tilde{\sigma}{}^a)^\alpha G_a+ {\cal O}(\vartheta^3)\, , \;
\end{aligned}
\ee
whose derivation, in its turn, requires to know $\tilde{L}^{\alpha\beta}$ up to the first order
\begin{eqnarray}\label{ssp-susy-L}
\tilde{L}^{\alpha\beta}(x)= \theta^{(\alpha}\epsilon^{\beta )} \bar{R}(x). \;  \qquad
\end{eqnarray}

To find the expressions for $\epsilon^M(x,\theta,\bar{\theta})$ one can first calculate the inverse supervielbein matrices, which up to the first order in the fermionic coordinates have the following form
\be
\begin{aligned}\label{EAM=}
E_a^{m }=& e_a^{m }  + \ii\psi_a\sigma^m\bar\theta - \ii\theta \sigma^m\bar\psi_a +{\cal O}(\vartheta^2)
\, ,  \\  E_a^{\mu }\delta_\mu^\alpha =& - \psi_a^\alpha+ \theta^\beta \hat{\omega}_\beta{}^\alpha
- 2\ii(\psi_a\sigma^b\bar\theta - \ii\theta \sigma^b\bar\psi_a)\psi_b^\alpha  \\ &  - \frac \ii{16}
[2\theta^\alpha G_a + (\theta \sigma_a\tilde\sigma_b)^{\alpha} G^d]- \frac \ii{16} (\bar\theta\tilde\sigma_a)^{\alpha} R
+{\cal O}(\vartheta^2)
\, ,   \\  E_\alpha^{m }=& \ii(\sigma^m\bar{\theta})_\alpha+{\cal O}(\vartheta^2)\; , \qquad  E_{\dot\alpha}{}^{m }= \ii({\theta}\sigma^{m})_{\dot\alpha} +{\cal O}(\vartheta^2)\; ,  \\ E_\alpha{}^{\nu }=&\delta _\alpha{}^{\nu } +{\cal O}(\vartheta^2)\; .
\end{aligned}
\ee
However, a significantly more economic way is to write
$$\iota_\epsilon E^{A}:=\iota_\epsilon E^{A}_{(0)}+\iota_\epsilon E^{A}_ {(1)}+\iota_\epsilon E^{A}_{(2)}+ ...=(\epsilon^{M}_{(0)} +\epsilon^{M}_{(1)} + \epsilon^{M}_{(2)}+...) ( E_{(0)M}^{~~~A}+  E^{~~~A}_{(1)M}+  E^{~~~A}_{(2)M}+ ...)\;.  $$
Then, by comparing the $\vartheta$-orders $(n)$ of the left and the right hand side one obtains the expressions for $ \epsilon^{M}_{(1)}, \dots \epsilon^{M }_{(4)}$.

In this way after some algebra we find that the WZ gauge is preserved by the superdiffeomorphisms with parameters
\be
\begin{aligned}\label{adiffWZ}
 \epsilon^{\alpha}(z)=& \epsilon^\alpha    +\ii\left(\theta \sigma^m\bar\epsilon-  \epsilon\sigma^m\bar\theta \right)\; \left[\psi_m^\alpha - \theta^\beta \hat{\omega}_{m\beta}{}^\alpha
 +i \left(\psi_m \sigma^n\bar\theta-  \theta\sigma^n\bar\psi_m \right)\psi_n^\alpha
 \right]  \\ & -
 \frac 1{16} \left(\theta \sigma^m\bar\epsilon-  \epsilon\sigma^m\bar\theta \right)\; \left[2
 \theta^\alpha G_m + (\theta \sigma_{mn})^{\alpha} G^n+2 (\bar\theta\tilde\sigma_a)^{\alpha} R(x)\right] +  {\cal O}(\vartheta^3)\, ,
\\
 \epsilon^m(z)=& -\ii \left(\theta \sigma^m\bar\epsilon-  \epsilon\sigma^m\bar\theta \right)\;
 \left(\delta_n{}^m - \ii \theta\sigma^m\bar\psi_n +  \ii\psi_n \sigma^m\bar\theta \right)
   + {\cal O}(\vartheta^3) \; ,
\end{aligned}
\ee
which at the linear order in $\theta$ simulate the rigid supersymmetry transformations of the flat superspace.

Equations \eqref{adiffWZ} allow one to determine the supersymmetry transformation of the goldstino,  eqs. (\ref{varTh}), and of the bosonic coordinates
\begin{subequations}
\begin{align}
\label{varTh=C}
 \delta \theta^{\alpha}=&- \epsilon^\alpha   -\ii\left(\theta \sigma^m\bar\epsilon-  \epsilon\sigma^m\bar\theta \right)\; \left[\psi_m^\alpha +\nabla_m \theta{}^\alpha
 -\ii \left[ \theta\sigma^n\bar\psi_m- \psi_m \sigma^n\bar\theta \right)(\psi_n^\alpha+\nabla_n \theta{}^\alpha)
 \right] \nn\\ & +
 \frac 1{16} \left[\theta \sigma^a\bar\epsilon-  \epsilon\sigma^a\bar\theta \right)\; \left(2
 \theta^\alpha G_a + (\theta \sigma_{ab})^{\alpha} G^b+2 (\bar\theta\tilde\sigma_a)^{\alpha} R\right] +  {\cal O}(\vartheta^3)\, ,
\\ \label{varX=C}
 \delta x^m=&\, \ii \left(\theta \sigma^n\bar\epsilon-  \epsilon\sigma^n\bar\theta \right)
 \left(\delta_n{}^m - \ii \theta\sigma^m\bar\psi_n +  \ii\psi_n \sigma^m\bar\theta \right)
   + {\cal O}(\vartheta^3) \; .
\end{align}
\end{subequations}
With some more efforts one can determine the form of the supersymmetry transformations to  all orders in $\vartheta$.

\section{The decomposition of the supervielbeins in the Majorana spinor representation}\label{Majorana}
\setcounter{equation}0
\def\theequation{\ref{Majorana}.\arabic{equation}}

For completeness, we give the generic form of the expansion of the  supervielbeins $E^A(x,\vartheta)$ in series of the Majorana spinor coordinates $\vartheta^{\hat\alpha}=(\theta^{\alpha},\bar\theta_{\dot\alpha})$ in the Wess-Zumino gauge obtained by adopting the same approach used in \cite{Grisaru:2000ij}.

The gamma-matrices are constructed with the Pauli matrices in a conventional way
\be\label{gamma}
\gamma^{a\,\hat\alpha}_{\hat\beta}=
  \left( {\begin{array}{cc}
   0 & \ii\sigma^a_{\alpha\dot\beta} \\       \ii\tilde\sigma^{a\dot\beta\alpha} & 0       \end{array} } \right)\,
\ee
and the spinor indices are lowered by the charge conjugation matrix
\be
C_{\hat\alpha\hat\beta}=
  \left( {\begin{array}{cc}
   \varepsilon_{\alpha\beta} & 0 \\   0     & \varepsilon^{\dot\alpha\dot\beta} \end{array} } \right)\,.
\ee
The torsion constraints are now chosen to have no $T_{bc}^a$ component and have the following form
\bea\label{Torsion}
T^a=&\frac 12 E\wedge \gamma^a E,\nn\\
T^{\hat\alpha}=&E^{\hat\beta}\wedge E^cT_{c\hat\beta}{}^{\hat\alpha}+\frac 12 E^b\wedge E^c T_{cb}^{\hat\alpha}\,,
\eea
where
\be\label{X}
16\,T_{a\hat\beta}{}^{\hat\alpha}= (\gamma_a\gamma_b\gamma_5)_{\hat\beta}{}^{\hat\alpha}G^b+ (\gamma_5)_{\hat\beta}{}^{\hat\alpha}G^b+ \large(\gamma_a(1+\gamma_5)\large)_{\hat\beta}{}^{\hat\alpha}R+ \large(\gamma_a(1-\gamma_5)\large)_{\hat\beta}{}^{\hat\alpha}\bar R\,.\nn
\ee
The superspace 2-form curvature obeys the following constraints
\bea\label{R}
R_{ab} &= & E^{\hat \alpha}\wedge( E \gamma_{[a}T_{b]})_{\hat \alpha} + E^{c}\wedge (E \gamma_{c}T_{ab} - \frac 32
E \gamma_{[a}T _{bc]})+\frac 12 E^c \wedge E^d\mathcal R_{dc, ab}\,.
\eea

Using the form of the constraints \eqref{Torsion}-\eqref{R} and their Bianchi identities \eqref{Bianchi} one gets the following $\vartheta$-expansion of the supervielbeins
\be\label{MEalpha}
E^{\hat\alpha}=\psi^{\hat\alpha}+E^{\hat\alpha}_{(1)}+E^{\hat\alpha}_{(2)}+E^{\hat\alpha}_{(3)}+E^{\hat\alpha}_{(4)}+ E^{\hat\alpha}_{(5)}\,,
\ee
\be\label{MEa}
E^{a}=e^{a}+\Big(\psi^{\hat\alpha}+\frac 12 E^{\hat\alpha}_{(1)}+\frac 13 E^{\hat\alpha}_{(2)}+\frac 14 E^{\hat\alpha}_{(3)}+\frac 15 E^{\hat\alpha}_{(4)}\Big)\gamma^a_{\hat\alpha\hat\beta}\vartheta^{\hat\beta}\,,
\ee
with
\be\label{defs1}
E^{\hat\alpha}_{(n)}=\frac 1{n!}\Big(\hat\nabla\,\vartheta^{\hat\gamma\,}V_{(n-1)}{}_{\hat\gamma}^{\hat\alpha}+\Sigma^{\hat\alpha}_{(n)}\Big)
\ee
and
\bea\label{defs2}
V_{(n-1)}{}_{\hat\gamma}^{\hat\alpha}=\vartheta^{\hat\beta_1}\hdots\vartheta^{\hat\beta_{n-1}}V^{\hat\alpha}_{[\hat\beta_1 \hdots\hat\beta_{n-1}]\hat\gamma}(x)\,,\qquad
\Sigma^{\hat\alpha}_{(n)}=\vartheta^{\hat\beta_1}\hdots\vartheta^{\hat\beta_n}\Sigma^{\hat\alpha}_{[\hat\beta_1 \hdots\hat\beta_n]}(x).
\eea
Hence, proceeding along the lines of \cite{Grisaru:2000ij}, we obtain
\be\label{sigma}
\begin{aligned}
E^{\hat\alpha}_{(1)}=\hat\nabla\vartheta^{\hat\gamma}V_{(0)}{}_{\hat\gamma}^{\hat\alpha}+\Sigma^{\hat\alpha}_{(1)}
=&\hat\nabla\vartheta^{\hat\alpha}+\vartheta^{\hat\beta}e^c(x)T_{c\hat\beta}^{\hat\alpha}\,,\\
2E^{\hat\alpha}_{(2)}=\hat\nabla\vartheta^{\hat\gamma}V_{(1)}{}_{\hat\gamma}^{\hat\alpha}+\Sigma^{\hat\alpha}_{(2)}=&\vartheta^{\hat\beta}\vartheta^{\hat\gamma}\Big(e^cH^{\hat\alpha}_{c\hat\gamma\hat\beta}
-\psi^{\hat\delta}K^{\hat\alpha}_{\hat\delta\hat\gamma\hat\beta}\Big)\,,\\
3!\,E^{\hat\alpha}_{(3)}=\hat\nabla\vartheta^{\hat\gamma}V_{(2)}{}_{\hat\gamma}^{\hat\alpha}+\Sigma^{\hat\alpha}_{(3)}
=&\hat\nabla\vartheta^{\hat\delta}\vartheta^{\hat\beta}\vartheta^{\hat\gamma}K_{\hat\delta\hat\gamma\hat\beta}^{\hat\alpha}\\
&+\vartheta^{\hat\beta}\vartheta^{\hat\gamma}\vartheta^{\hat\delta}\Big(e^cT^{\hat\sigma}_{c\hat\delta}K^{\hat\alpha}_{\hat\sigma\hat\gamma\hat\beta}
+e^c\mathcal D_{\hat\delta}H^{\hat\alpha}_{c\hat\gamma\hat\beta}
-\psi^{\hat\sigma}(\gamma^b_{\hat\sigma\hat\delta} H^{\hat\alpha}_{b\hat\gamma\hat\beta}+\mathcal D_{\hat\delta}K^{\hat\alpha}_{\hat\sigma\hat\gamma\hat\beta})\Big)\,,\\
4!\,E^{\hat\alpha}_{(4)}=\hat\nabla\vartheta^{\hat\gamma}V_{(3)}{}_{\hat\gamma}^{\hat\alpha}+\Sigma^{\hat\alpha}_{(4)}
=& \hat\nabla\vartheta^{\hat\tau}\vartheta^{\hat\beta}\vartheta^{\hat\gamma}\vartheta^{\hat\delta}\Big(
2D_{\hat\delta}K^{\hat\alpha}_{\hat\tau\hat\gamma\hat\beta}+\gamma^c_{\hat\delta\hat\tau}H^{\hat\alpha}_{c\hat\gamma\hat\beta}\Big)
\\
&+\vartheta^{\hat\beta}\vartheta^{\hat\gamma}\vartheta^{\hat\delta}\vartheta^{\hat\tau}\Big[e^cT^{\hat\sigma}_{c\hat\tau}(2D_{\hat\delta}K^{\hat\alpha}_{\hat\sigma\hat\gamma\hat\beta}+\gamma^c_{\hat\delta\hat\sigma}H^{\hat\alpha}_{c\hat\gamma\hat\beta})\\
&+e^c(\mathcal D_{\hat\tau}\mathcal D_{\hat\delta} H^{\hat\alpha}_{c\hat\gamma\hat\beta}
+H^{\hat\sigma}_{c\hat\tau\hat\delta}K^{\hat\alpha}_{\hat\sigma\hat\gamma\hat\beta})\nn\\
&+\psi^{\hat\sigma}(\mathcal D_{\hat\tau}\mathcal D_{\hat\delta} K^{\hat\alpha}_{\hat\sigma\hat\gamma\hat\beta}
+2\gamma^c_{\hat\sigma\hat\tau}\mathcal D_{\hat\delta}H^{\hat\alpha}_{c\hat\gamma\hat\beta}-K^{\hat\rho}_{\hat\sigma\hat\tau\hat\delta}K^{\hat\alpha}_{\hat\rho\hat\gamma\hat\beta})
\Big]\,,
\end{aligned}
\ee
where $H$ and $K$ are expressed in terms of components of the curvature and torsion as follows
\be\label{H+K}
H^{\hat\alpha}_{c\hat\gamma\hat\beta}=\mathcal R_{c[\hat\gamma\hat\beta]}{}^{\hat\alpha}+\mathcal D_{[\hat\gamma}T_{\hat\beta]c}^{\alpha}\,,\qquad
K_{\hat\delta\hat\gamma\hat\beta}^{\hat\alpha}=\mathcal R_{\hat\delta[\hat\gamma\hat\beta]}{}^{\hat\alpha}+\gamma^c_{\hat\delta[\hat\gamma}T_{\hat\beta]c}^{\alpha}\,.
\ee



\if{}
\bibliographystyle{abe}
\bibliography{references}{}
\end{document}